\documentclass[12pt]{iopart}
\usepackage{epsfig}

\newcommand{\pprime}{{\prime\prime}}

\newcommand{\bra}{\langle}
\newcommand{\ket}{\rangle}
\newcommand{\lav}{\left\langle}
\newcommand{\rav}{\right\rangle}
\newcommand{\order}{{\mathcal O}}

\newcommand{\be}{\begin{equation}}
\newcommand{\ee}{\end{equation}}
\newcommand{\bea}{\begin{eqnarray}}
\newcommand{\eea}{\end{eqnarray}}
\newcommand{\beastar}{\begin{eqnarray*}}
\newcommand{\eeastar}{\end{eqnarray*}}
\newcommand{\bd}{\begin{displaymath}}
\newcommand{\ed}{\end{displaymath}}

\newcommand{\vsp}{\vspace*{3mm}}

\newcommand{\ie}{{\it i.e.}}
\newcommand{\eg}{{\it e.g.}}

\newcommand{\eq}[1]{~(\ref{#1})}

\newcommand{\N}{{\rm I\!N}}

\newcommand{\bc}{\ensuremath{\mathbf{c}}}

\newcommand{\bk}{\ensuremath{\mathbf{k}}}

\newcommand{\bomega}{{\mbox{\boldmath $\omega$}}}

\newcommand{\ci}{{\bf c}}
\newcommand{\kav}{{\langle k\rangle}}

\newcommand{\Partition}{\mathcal{Z}}
\newcommand{\Pcode}{\mathcal{P}[{\rm conn}|{\bf c},k,k^\prime]}
\newcommand{\Pco}{\mathcal{P}[{\rm conn}|{\bf c}]}
\newcommand{\Proba}{{\rm Prob}}
\newcommand{\ensemble}{{\sum_{\bf c} \Proba({\bf c}|p,Q)}}

\newcommand{\Qij}{Q(k_i,k_j)}

\newcommand{\Qk}{Q(k,k^\prime)}

\newcommand{\Pik}{\Pi(k,k^\prime)}

\newcommand{\Pitildec}{\tilde{\Pi}(k,k^\prime|\ci)}
\newcommand{\Pitildeco}{\tilde{\Pi}(k,k^\prime|\ci_0)}

\newcommand{\Fkq}{F(k|Q)}
\newcommand{\Fkqp}{F(k^\prime|Q)}
\newcommand{\Pic}{\Pi(k,k^\prime|\ci)}
\newcommand{\Pico}{\Pi(k,k^\prime|\ci_0)}
\newcommand{\de}{\delta_{c_{ij},1}}
\newcommand{\deO}{\delta_{c_{ij},0}}
\newcommand{\deki}{\delta_{k_i,k_i({\bf c})}}
\newcommand{\dek}{\delta_{k,k_i({\bf c})}}
\newcommand{\dekp}{\delta_{k^\prime,k_j({\bf c})}}
\newcommand{\Prob}{{\rm prob}}

\begin{document}\small

\title[Tailored graph ensembles as proxies or null models for real networks]
{Tailored graph ensembles as proxies or null models for real networks I: tools for quantifying structure}

\author{A Annibale$^\dag$, ACC Coolen$^{\dag\ddag}$, LP Fernandes$^\ddag$,\\ F Fraternali$^\ddag$ and J Kleinjung$^\S$}
\address{$\dag$ ~ Department of Mathematics, King's College London, The Strand,
London WC2R 2LS, United Kingdom}
\address{$\ddag$ Randall Division of Cell and Molecular Biophysics,
King's College London, New Hunt's House, London SE1 1UL, United Kingdom}
\address{$\S$~ Medical Research,
 The Ridgeway, Mill Hill,
  London NW7 1AA, United Kingdom}

\begin{abstract}
We study the tailoring of structured random graph ensembles to real networks,
with the objective of generating precise and practical mathematical
tools for quantifying and comparing network topologies macroscopically,
beyond the level of degree statistics. Our family of ensembles can produce graphs
 with any prescribed degree distribution and any degree-degree correlation function,
its control parameters can be calculated fully analytically, and as a result
we can calculate (asymptotically) formulae for entropies and complexities, and for
information-theoretic distances between networks,
 expressed directly and explicitly in terms of their
measured degree distribution and degree correlations.
\end{abstract}

\ead{alessia.annibale@kcl.ac.uk,ton.coolen@kcl.ac.uk}

\section{Introduction}

In the study of natural or synthetic signaling networks, one of the key questions is how network structure
relates to the execution of the process which it supports. This is especially true in
systems biology, where, for instance, our understanding of how the structure of
protein-protein interaction networks (PPIN) relates to their biological functionality is vital in the design of a new generation of
intelligent and personalized  medical interventions.
In recent years, high-throughput proteomics has allowed for the drafting of large PPIN
data sets, for different organisms, and with different experimental techniques and degrees of accuracy.
With this accumulation of information, we now face the challenge of
analyzing these data from a complex networks perspective, and using them optimally in order to increase
our understanding of how PPIN control the functioning of cells, both in healthy and in diseased conditions.
A prerequisite for achieving this is the availability of precise mathematical tools with which to
quantify topological structure in large observed networks, to compare
network instances and distinguish between meaningful and `random' structural features.
These tools have to be both systematic, i.e. with a sound statistical or information-theoretic basis, but also practical, i.e. preferably formulated in terms of explicit formulae as opposed to tedious numerical simulations.

Many quantities have been proposed for characterizing the
structure of networks, such as degree distributions \cite{AlbBar02}, degree
sequences \cite{BarAlb99},
degree correlations \cite{PasVazVes01}
and assortativity \cite{Newman02}, clustering coefficients \cite{WatStr98}, and
community structures \cite{NewLei07}.
To assess the relevance of an observed topological feature in a network,
a common strategy is to compare it against similar observations in so-called `null models',
defined as randomized versions of the original network which retain some
features of the original one.
The choice of which topological features to conserve in the randomized models
was mostly limited to degree distributions and degree sequences.
Such null models were used to assess the statistical relevance of network
motifs in real networks, viz. patterns which were observed significantly more often in the real networks
than in their randomized counterparts \cite{MasSne02,MasSneZal04,SheMilMonAlo02}.
Whether any such proposed motif is indeed functionally important and/or represent (evolutionary) arisen principles,
is however not obvious; topological deviations from randomized networks could also be
merely irrelevant consequences of some neglected structural property of the network,
i.e. the result of an inappropriate null hypothesis rather than of a
distinctive feature
of the process \cite{BjoSch08,Artzy04}.
The definition and generation of good
null models for benchmarking topological measures of real world graphs (and the dynamical processes
which they enable) is a nontrivial issue. Similarly, in comparing observed networks (which, as a result of experimental noise, will usually
not even have identical nodes), one would seek to focus on the values of macroscopic topological observables, and know the typical properties of networks with the observed features.

In recent years there have been
efforts to define
and generate random graphs whose topological features can be controlled
 and tailored to experimentally observed networks.
In \cite{PerCoo08} a parametrized random graph ensemble was defined where graphs have a prescribed
degree sequence, and links are drawn in a way that allows for
preferential attachment on the basis of arbitrary two-degree kernels.
In this paper we generalize the definition of this ensemble, and show that it
can be tailored asymptotically to the generation of graphs with any prescribed degree
distribution and any prescribed degree correlation function (and that it is a maximum entropy ensemble, given the
degree correlations).
Moreover, in spite of its parameter space being in principle infinitely large,
in contrast to most random graph ensembles used to mimic real networks,
we can derive explicit analytical formulae for the parameters of the ensemble, to leading order in system size,
expressed directly in terms of the observed characteristics of the network given.
Graphs from this ensemble are thus ideally suited to be used as
either proxies or null models for observed networks, depending on the  question to be answered.

Statistical mechanics approaches
have been proposed to quantify the information content
of network structures. Especially the (Shannon or Boltzmann) entropy has been instrumental in characterizing the complexity of network
ensembles \cite{Bianconi08,BiaCooPer08,Bianconi09}.
Here, the crucial availability of analytical expressions for the
parameters of our ensemble will enable us to derive explicit formulae, in the thermodynamic
limit (based on combinatorial and saddle-point arguments),
for our ensemble's Shannon entropy, and hence also for the complexity of its typical graphs.
These formulae are compact and transparent, and expressed solely and explicitly in terms of the degree distribution
and the degree correlations that our ensemble is targeting.
Finally, along similar lines we can obtain an information theoretic distance between networks, again
expressed solely in terms of their degree distributions and degree correlations.
A companion paper \cite{Fernandes} will be devoted to large scale applications to PPIN data of these complexity and distance measures;
here we focus on their mathematical derivation. Although there is no need for numerical sampling in our derivations
(all results can be obtained analytically), we note that exact algorithms for generating random graphs from  the proposed ensemble exist \cite{CooDemAnn09}.

\section{Definitions and properties of network topology characterizations}
\label{Sec:definitions}

\subsection{Networks, degree distributions, and degree correlation functions}

We study networks (or graphs) of $N$ nodes (or vertices), labeled
by Roman indices $i,j,\ldots$ etc, where every vertex can be connected
to other vertices by undirected links (or `edges').
The microscopic structure of such a network is defined in full by
an $N\times N$ matrix of binary variables $c_{ij} \in \{0,1\}$, where the nodes
$i$ and $j$ are connected by a link if and only if $c_{ij}=1$.
We define $c_{ij}=c_{ji}$ and $c_{ii}=0$ for all $(i,j)$, and we
abbreviate $\ci=\{c_{ij}\}$. Henceforth, unless indicated otherwise, any summation over Roman indices will always run
over the set $\{1,\ldots,N\}$.

A standard way of characterizing the topology of a network $\ci$,
as e.g. observed in a biological or physical system under study, is to measure for each vertex $i$ the degree
$k_i(\ci)=\sum_{j} c_{ij}$, the number of the links to this vertex. From these numbers then follow
the empirical degree distribution $p(k|\ci)$ and the observed average connectivity $\overline{k}(\ci)$:
\be
p(k|\ci)=\frac{1}{N}\sum_{i} \delta_{k,k_i(\ci)},~~~~~~~~\overline{k}(\ci)=\frac{1}{N}\sum_i k_i(\bc)
\label{eq:pk}
\ee
(using the Kronecker $\delta$-symbol for $n,m\in\N$, defined as $\delta_{nm}=1$ for $n\!=\!m$ and $\delta_{nm}=0$ otherwise).
Objects such as $p(k|\ci)$ have the advantage of being macroscopic in nature, allowing for size-independent characterization of network topologies, and for comparing networks that differ in size.
However, networks with the same degree distribution can still differ profoundly
in their microscopic structures. We need observables that capture additional
topological information, in order to discriminate between different networks
with the same degree distribution (\ref{eq:pk}).

To construct macroscopic observables that quantify network topology beyond the level of degree statistics, it is natural
to consider how the likelihood for two nodes of a network $\ci$ to be connected
depends on their degrees, which is measured by the degree correlation function

\be
\Pitildec=\frac{\Pcode}{\Pco}
\label{Pitildecdef}
\ee
Here $\Pcode$ is the probability for two randomly drawn nodes with
degrees $(k,k^\prime)$ to be connected, and $\Pco$ is the overall probability
for two randomly drawn nodes to be connected, irrespectively of their
degrees, viz.
\begin{eqnarray}
{\mathcal P}[{\rm conn}|\bc,k,k^\prime]&=& \frac{\sum_{i\neq j} c_{ij}~\delta_{k,k_i(\bc)}\delta_{k^\prime,k_j(\bc)}}
{ \sum_{i\neq j}\delta_{k,k_i(\bc)}\delta_{k^\prime,k_j(\bc)}}
\label{eq:Pcode}
\\
{\mathcal P}[{\rm conn}|\bc]&=&  \frac{1}{N(N\!-\!1)}\sum_{i\neq j} c_{ij}~=\frac{\overline{k}(\bc)}{N\!-\!1}
\label{eq:Pco}
\end{eqnarray}
By definition, $\Pitildec$ is symmetric under
exchanging $k$ and $k^\prime$.
For simple networks $\ci_0$, with some degree distribution $p(k)$ but without any micro-structure beyond that required by
$p(k)$\footnote{In section 2 we will give a precise and more general mathematical definition of `simple networks', relative to some imposed macroscopic feature such as the degree distribution $p(k)$.}, it is known (see e.g. \cite{DorogovtsevMendesBook} and references therein) that in the limit $N\to\infty$ one finds
\be
\Pitildeco=kk^\prime\!/~\overline{k}^2(\bc).
\label{eq:Pitildeco}
\ee
It follows that those topological properties of a given (large) network $\ci$, that manifest themselves at the level of degree correlations and
cannot be attributed simply to its degree statistics,
can be quantified by a deviation from the simple law\eq{eq:Pitildeco}; see also \cite{MasSne02,IvaWalRei08a,IvaWalRei08b}.
One is therefore led in a natural way to the introduction of the {\em relative} degree correlations
\begin{eqnarray}
\Pic&=&\frac{\Pitildec}{\Pitildeco}
=\frac{\Pcode}{\Pco}\frac{\overline{k}^2(\bc)}{kk^\prime}.
\label{eq:Picdef}
\end{eqnarray}
By definition, $\Pi(k,k^\prime|\ci_0)=1$ for sufficiently large simple networks $\ci_0$, whereas any statistically relevant deviation from $\Pic=1$
signals the presence in network $\ci$ of underlying criteria for
connecting nodes beyond its degrees.
Just like $p(k|\ci)$, $\Pic$ is again a {\it macroscopic} observable that can be measured
directly and at low computation cost. It is therefore a natural tool for quantifying and comparing network
structures beyond the level of degree statistics.

\subsection{Properties of the relative degree correlation function}

 To prepare the ground for proving some asymptotic mathematical properties of the relative degree correlation function $\Pic$, we first simplify the denominator of (\ref{eq:Pcode}):
\begin{eqnarray}
\sum_{i\neq j}\delta_{k,k_i(\bc)}\delta_{k^\prime,k_j(\bc)}&=& \sum_{ij}\delta_{k,k_i(\bc)}\delta_{k^\prime,k_j(\bc)}-\delta_{kk^\prime}\sum_i\delta_{k,k_i(\bc)}
\nonumber
\\
&=& N^2\big[p(k|\ci)p(k^\prime|\ci)-N^{-1}\delta_{kk^\prime}p(k|\ci)\big]
\end{eqnarray}
Upon inserting the result for (\ref{eq:Pco}) together with (\ref{eq:Pcode}) into (\ref{eq:Picdef})
we then find that
\begin{eqnarray}
\Pic&=&\frac{N^{-1}\sum_{i\neq j} c_{ij} \delta_{k,k_i(\bc)}\delta_{k^\prime,k_j(\bc)}}
{[p(k|\ci)-N^{-1}\delta_{kk^\prime}]p(k^\prime|\ci)}
~\frac{\overline{k}(\bc)}{kk^\prime}~(1\!-\!N^{-1})
\label{eq:Picdef2}
\end{eqnarray}
and hence, using $c_{ii}=0$ for all $i$,
\begin{eqnarray}
\lim_{N\to\infty}
\Pic=&\lim_{N\to\infty}\frac{\overline{k}(\bc) \sum_{ij}c_{ij}\dek\dekp}{Np(k|\ci)p(k^\prime|\ci)kk^\prime}.
\label{eq:Pic}
\end{eqnarray}
We are now in a position to establish three identities obeyed by $\Pic$. The first two of these, viz. (\ref{eq:selfPic},\ref{eq:Pinorm}), are the main ones; they are used frequently in mathematical manipulations of subsequent sections. The third provides the physical intuition behind (\ref{eq:selfPic},\ref{eq:Pinorm}). It is assumed implicitly in all proofs that $\overline{k}(\bc)$ remains finite for $N\to\infty$ and that the  limits $N\to\infty$ exist.
\begin{itemize}
\item{\em Linear constraints:}
\bea
\forall k\in\N:&~~~&
\lim_{N\to\infty}
\sum_{k^\prime} \frac{k^\prime p(k^\prime|\ci)}{\overline{k}(\bc)}\Pic=1
\label{eq:selfPic}
\end{eqnarray}
These are easily verified for simple graphs $\ci_0$, for which
$\Pico=1~\forall k, k^\prime$. However, they turn out to hold
 for {\it any} graph $\ci$, as can be proven using (\ref{eq:Pic}) as follows:
\begin{eqnarray}
\lim_{N\to\infty}
\sum_{k^\prime} \frac{k^\prime p(k^\prime|\ci)}{\overline{k}(\bc)}\Pic
&=&
\lim_{N\to\infty}\frac{\sum_{k^\prime}\sum_{ij}c_{ij}\dek\dekp}{Np(k|\ci)k}
\nonumber
\\
&=&\lim_{N\to\infty}\frac{\sum_{i}k_i(\ci)\dek}{Np(k|\ci)k}=1.
\eea
\item
{\em Normalization:}
\bea
\lim_{N\to\infty}
\sum_{kk^\prime} \frac{k p(k|\ci)}{\overline{k}(\bc)}\frac{k^\prime p(k^\prime|\ci)}{\overline{k}(\bc)}\Pic=1
\label{eq:Pinorm}
\end{eqnarray}
This follows directly from (\ref{eq:selfPic}) upon multiplying both sides by $kp(k)/\bra k\ket$, followed by summation over all $k$.
\item
{\em Interpretation of the linear constraints:}
\bea
\lim_{N\to\infty}\sum_{k^\prime} \frac{k^\prime p(k^\prime|\ci)}{\overline{k}(\bc)}\Pic&=& \lim_{N\to\infty}\frac{N}{k}\mathcal{P}[{\rm conn}|{\bf c},k]
\label{eq:LHSself}
\eea
Here
 $\mathcal{P}[{\rm conn}|{\bf c},k]$ is defined as
the probability that two randomly
drawn nodes, one having degree $k$, are connected:
\begin{eqnarray}
\mathcal{P}[{\rm conn}|{\bf c},k]&=&\frac{\sum_{i\neq j}c_{ij}\delta_{k,k_i(\bc)}}{\sum_{i\neq j}\delta_{k,k_i(\bc)}}
=\frac{\sum_{ij}c_{ij}\delta_{k,k_i(\bc)}}{N^2p(k|\ci)}~(1+\order(N^{-1}))
\end{eqnarray}
 The proof is elementary:
\bea
\hspace*{-15mm}
\lim_{N\to\infty}\sum_{k^\prime} \frac{k^\prime p(k^\prime|\ci)}{\overline{k}(\bc)}\Pic&=&
\lim_{N\to\infty} \frac{N}{k}\frac{\sum_{ij}c_{ij}\dek}{N^2 p(k|\ci)}=\lim_{N\to\infty}\frac{N}{k}\mathcal{P}[{\rm conn}|{\bf c},k]
\eea
We conclude that our first (proven) identity (\ref{eq:selfPic}) boils down to the claim that for large $N$ one has
$\mathcal{P}[{\rm conn}|{\bf c},k]=k/N$ (modulo irrelevant orders in $N$), which is easily understood.
\end{itemize}
We end this section with two further observations.
First, the relations (\ref{eq:selfPic}) involve the degree distribution, so one must expect
that the possible values for $\Pic$
are dependent upon (or constrained by)
$p(k|\ci)$. Second, several other useful properties of the kernel $\Pic$ can be extracted from  (\ref{eq:selfPic}).
For instance, the only separable kernel $\Pi$
is $\Pic=1$ for all $(k,k^\prime)$: a separable kernel is of the form $\Pic=G(k|\ci)G(k^\prime|\ci)$ for some function $G(k|\ci)$ ($\Pi$ being symmetric), and insertion of this form into (\ref{eq:selfPic}) leads immediately to $G(k|\bc)=1$ for all $k,\ci$.

\section{Random graphs with controlled macroscopic structure }

\subsection{Definition of the random graph ensembles}

To study the signalling properties of real-world networks,
 or generate `null models' to assess the relevance of observed topological features, one needs random
graph ensembles in which one can control the topological characteristics one is interested in
and `tune' these to match the characteristics of the observed networks.
Most ensembles studied in literature so far have focused on producing graphs with controlled degree statistics.
The suggestion that (\ref{eq:Picdef}) can be used for {\em identifying} network complexity beyond degree statistics
goes back at least to \cite{MasSne02,DorogovtsevMendesBook,IvaWalRei08a,IvaWalRei08b}. In contrast to these earlier studies,
which were mostly limited to measuring (\ref{eq:Picdef}) for real networks, here we take further mathematical steps
that will allow us to use (\ref{eq:Picdef}) as a systematic tool for
{\em quantifying} complexity and distances in network structure beyond degree statistics.
This requires generating random graphs in which we can control at will both the degree distribution $p(k)$ and the relative degree correlations $\Pi(k,k^\prime)$.

It will turn out that we can achieve our objectives with the following random graph ensembles,
in which all degrees $k_i$ are drawn randomly and independently from $p(k)$, and where in addition the edges are
drawn in a way that allows for preferential attachment on the basis of an
arbitrary symmetric function $Q(k,k^\prime)$ of the degrees of the two vertices concerned:
\begin{eqnarray}
\hspace*{-15mm}
\Proba(\bc|p,Q)&=&\sum_{\bk}\Proba(\bc|Q,\bk)\prod_i p(k_i)
\label{eq:ensemble}
\\
\hspace*{-15mm}
\Proba({\bf c}|\bk,Q)&=&\frac{1}{\Partition(\bk,Q)}\prod_{i<j}\left[
\frac{\overline{k}}{N}\Qij\de+\left(1\!-\!\frac{\overline{k}}{N}\Qij\right)\!\deO
\right]\prod_i\deki
\label{eq:micro_ensemble}
\end{eqnarray}
Here $\Partition(\bk,Q)$ is a normalization constant
that ensures $\sum_{\bc}\Proba({\bf c}|\bk,Q)=1$ for all $(\bk,Q)$,
$\overline{k}=N^{-1}\sum_i k_i$,
and the function $Q$ must obey
 $\Qk\ge 0$ for all $(k,k^\prime)$ and $N^{-2}\sum_{ij}Q(k_i,k_j)=1$.
 The ensemble (\ref{eq:micro_ensemble}) with prescribed degrees $\bk=(k_1,...,k_N)$ was
defined and studied in
\cite{PerCoo08,PerCoo09}. We note that in the above ensemble one will have $\overline{k}=\sum_k p(k)k+\order(N^{-1/2})$.

Upon making the simplest choice  $\Qk=1$ for all $(k, k^\prime)$
one retrieves from (\ref{eq:ensemble}) the `flat' ensemble, where once the individual degrees are drawn randomly from $p(k)$, all  graphs $\ci$ with
the prescribed degrees carry equal probability:
\begin{eqnarray}
\Proba(\bc)&=& \sum_{\bk}\Big[\prod_i p(k_i)\Big]
 \frac{
\prod_i\delta_{k_i,k_i(\bc)}}
{\sum_{\bc^\prime}
\prod_i\delta_{k_i,k_i(\bc^\prime)}}.
\label{eq:flat}
\end{eqnarray}
This follows from the property that for $\Qk=1$ the factor $\prod_{i<j}[\ldots\delta_{c_{ij},1}+\ldots\delta_{c_{ij},0}]$
in (\ref{eq:ensemble}) depends on $\bc$ via the degrees $\{k_i(\bc)\}$ only, and will consequently drop out of the measure (\ref{eq:micro_ensemble}):
\begin{eqnarray}
\hspace*{-10mm}
\prod_{i<j}\Big[\frac{\bra k\ket}{N}\delta_{c_{ij},1}+(1-\frac{\bra k\ket}{N})\delta_{c_{ij},0}\Big]&=& \Big(1-\frac{\overline{k}}{N}\Big)^{\frac{1}{2}N(N-1)}e^{\sum_{i<j}c_{ij}\log\big[\frac{\overline{k}}{N}(1-\frac{\overline{k}}{N})^{-1}\big]}
\nonumber
\\
&=& \Big(1-\frac{\overline{k}}{N}\Big)^{\frac{1}{2}N(N-1)}e^{\frac{1}{2}N\overline{k}\log\big[\frac{\overline{k}}{N}(1-\frac{\overline{k}}{N})^{-1}\big]}
\end{eqnarray}


\subsection{Asymptotic properties of the ensembles}

One should expect
that macroscopic physical observables
such as $p(k|\bc)$ (\ref{eq:pk}) and $\Pi(k,k^\prime|\bc)$ (\ref{eq:Picdef2}) are self-averaging, and can therefore
be calculated, to leading
order in $N$, in terms of their expectation values over the ensemble (\ref{eq:ensemble})\footnote{Proving this self-averaging property explicitly
for the ensemble (\ref{eq:ensemble}) is trivial in the case of $p(k|\bc)$, and nontrivial but feasible in the case of $\Pi(k,k^\prime|\bc)$}.
We should therefore find that each graph drawn from (\ref{eq:ensemble})
 will for sufficiently large $N$ have as its degree distribution $p(k)$ and will have relative degree correlations identical to
\begin{eqnarray}
\Pik&=&\lim_{N\to\infty}
\ensemble\Pic\nonumber
\\
&=&\frac{\kav}{p(k)p(k^\prime)kk^\prime}\lim_{N\to\infty}\frac{1}{N}\ensemble\sum_{ij}c_{ij}\dek\dekp
\label{eq:Pikdef}
\end{eqnarray}
It turns out that (\ref{eq:Pikdef}) can be calculated analytically, and expressed in terms of $p(k)$ and $Q(k,k^\prime)$.
The first published result related to this connection in an appendix of \cite{PerCoo08} was unfortunately subject to an error; see the corrigendum
\cite{PerCoo09} for the correct relation as given below, of which the actual derivation is given in \ref{app:Pi_in_Q} of this present paper:
\be
\Pik=\frac{\Qk}{\Fkq\Fkqp}
\label{eq:Pik}
\ee
where the function $\Fkq$ is calculated self-consistently, for any $\Qk$, as the solution of
\begin{eqnarray}
\forall k:&~~~&\Fkq=\frac{1}{\kav}\sum_{k^\prime}p(k^\prime)k^\prime\frac{\Qk}{\Fkqp}.
\label{eq:selfF}
\end{eqnarray}
It is satisfactory to observe, upon eliminating $\Qk$ from (\ref{eq:selfF}) via (\ref{eq:Pik}), that (\ref{eq:selfF}) becomes identical to the
set of relations (\ref{eq:selfPic}) that we derived earlier for $\Pik$ solely on the basis of the latter's microscopic definition.
Clearly, (\ref{eq:selfPic}) must indeed hold for   every single graph of the
ensemble (\ref{eq:ensemble}), provided $N$ is sufficiently large.
On the other hand, for finite $N$ a typical graph of the ensemble (\ref{eq:ensemble})
will display deviations from (\ref{eq:Pik}) that are at least
of order $\order{(N^{-1})}$ (the difference between definition (\ref{eq:Picdef2})
and its asymptotic form (\ref{eq:Pic})), but possibly of order $\order{(N^{-1/2})}$
(the typical finite size corrections in empirical averages over $\order(N)$ independent samples).

Expression (\ref{eq:Pik}) also provides {\em en passant} the explicit proof that for graphs in which the only structure is that imposed by the degree sequence, viz.
those generated from (\ref{eq:flat}) corresponding to $Q(k,k^\prime)=1$ for all $(k,k^\prime)$, one indeed finds $\Pik=1$ for $N\to\infty$. Upon inserting $Q(k,k^\prime)=1$ into condition
(\ref{eq:selfF})  we find that $F^2(k)=1$ for all $k$, upon which  the desired result follows directly from (\ref{eq:Pik}).
\vsp

Asymptotically (i.e. in leading relevant orders in $N$), the probabilities (\ref{eq:ensemble}) to find graphs $\bc$ with the correct degree statistics, i.e. with degrees drawn randomly from $p(k)$, depends on $\bc$ via the degree distribution $p(k|\bc)$ and the kernel  $\Pi(k,k^\prime|\bc)$ only.
To see  this we study the following function for large $N$,
\begin{eqnarray}
\hspace*{-10mm}\Omega(\ci|p,Q)&=&-N^{-1}\log \Proba(\ci|p,Q)
=-N^{-1}\log \sum_{\bk}\prod_i \Big[p(k_i)\delta_{k_i,k_i(\bc)}\Big]e^{-N\Omega(\ci|\bk,Q)}
\nonumber
\\
\hspace*{-10mm}
&=& -\frac{1}{N}\sum_i \log p(k_i(\bc))+\Omega(\bc|\bk(\bc),Q)
\label{eq:Omega_def}
\end{eqnarray}
The leading order in
$\Omega(\ci|\bk,Q)=-N^{-1}\log p(\bc|\bk,Q)$ was studied in \cite{PerCoo08}. If $\bk(\bc)\neq \bk$ one has $\Omega(\bc|\bk,Q)=\infty$ (the degrees are imposed as strict constraints), whereas for
$\bk=\bk(\bc)$ one has
\begin{eqnarray}
\Omega(\ci|\bk,Q)&=&
\frac{1}{2}\overline{k}\log N+
\frac{1}{2}\overline{k}[\log \overline{k} -1] -N^{-1}\sum_i\log k_i!
+N^{-1}\sum_i k_i \log F(k_i|Q)
\nonumber
\\
&& - N^{-1}\sum_{i<j}c_{ij}\log Q(k_i,k_j)+\order(N^{-1})
\nonumber
\\
&=& \frac{1}{2}\overline{k}\log N+
\frac{1}{2}\overline{k}[\log \overline{k} -1] -N^{-1}\sum_i\log k_i!
+N^{-1}\sum_i k_i \log F(k_i|Q)
\nonumber
\\
&& - \frac{1}{2}\sum_{kk^\prime}\log Q(k,k^\prime)\frac{1}{N}\sum_{ij}c_{ij}\delta_{k,k_i}\delta_{k^\prime,k_j}+\order(N^{-1})
\label{eq:appen1_intermediate}
\end{eqnarray}
where $\overline{k}=N^{-1}\sum_i k_i$. We introduce the further short-hand $\tilde{p}(k)=N^{-1}\sum_i \delta_{k,k_i}$,
as well as the notation ${\sl o}(1)$ to denote finite size corrections that obey $\lim_{N\to\infty}{\sl o}(1)=0$
(to determine the exact scaling with $N$ of these corrections we would have to inspect e.g. the finite size corrections to
 (\ref{eq:Pik})).
We write the leading orders of (\ref{eq:appen1_intermediate}) in terms of the kernel $\Pi(k,k^\prime|\bc)$, using (\ref{eq:Pic}) and (\ref{eq:Pik}), and
substituting into (\ref{eq:Pik}) the present degree distribution $\tilde{p}(k)$, and find
\begin{eqnarray}
\Omega(\ci|\bk,Q)&=&
\frac{1}{2}\overline{k}\log N+
\frac{1}{2}\overline{k}[\log \overline{k} -1] -\sum_k \tilde{p}(k)\log k!
+\sum_k \tilde{p}(k)k \log F(k|Q)
\nonumber
\\
&& - \frac{1}{2}\sum_{kk^\prime}\log Q(k,k^\prime)N^{-1}\sum_{ij}c_{ij}\delta_{k,k_i(\bc)}\delta_{k^\prime,k_j(\bc)}+{\sl o}(1)
\nonumber
\\
&=&
\frac{1}{2}\overline{k}\log N+
\frac{1}{2}\overline{k}[\log\overline{k} -1] -\sum_k \tilde{p}(k)\log k!
+\sum_k \tilde{p}(k)k \log F(k|Q)
\nonumber
\\
&& - \sum_{kk^\prime}\frac{\tilde{p}(k)\tilde{p}(k^\prime)kk^\prime}{2\overline{k}}\Pic \log[\Pik \Fkq\Fkqp]
+{\sl o}(1)
\nonumber
\\
&=&
\frac{1}{2}\overline{k}\log N+
\frac{1}{2}\overline{k}[\log \overline{k} -1] -\sum_k \tilde{p}(k)\log k!
\nonumber
\\
&&
+\sum_k \tilde{p}(k)k \log F(k|Q)\Big[1-
\sum_{k^\prime}\frac{\tilde{p}(k^\prime)k^\prime}{\overline{k}}\Pic\Big]
\nonumber
\\&&
 - \sum_{kk^\prime}\frac{\tilde{p}(k)\tilde{p}(k^\prime)kk^\prime}{2\overline{k}}\Pic \log\Pik
+{\sl o}(1)
\nonumber
\\
&=&
\frac{1}{2}\overline{k}\log N+
\frac{1}{2}\overline{k}[\log \overline{k} -1] -\sum_k \tilde{p}(k)\log k!
\nonumber
\\
&&
 - \sum_{kk^\prime}\frac{\tilde{p}(k)\tilde{p}(k^\prime)kk^\prime}{2\overline{k}}\Pic \log\Pik
+{\sl o}(1)
\end{eqnarray}
where in the last step we used the identities (\ref{eq:selfPic}).
It subsequently follows (\ref{eq:Omega_def}) as
\begin{eqnarray}
\hspace*{-10mm}
\Omega(\ci|p,Q)
&=&
\frac{1}{2}\overline{k}(\bc)\log N
+
\frac{1}{2}\overline{k}(\bc)[\log \overline{k}(\bc) -1] -\sum_k p(k|\bc)\log k!
\\
&&
-\Omega[p(\bc),\Pi(\bc);p,\Pi]+{\sl o}(1)
\label{eq:asymptotic_ensemble}
\\[1mm]
\hspace*{-10mm}
\Omega[p(\bc),\Pi(\bc);p,\Pi]&=&
\sum_{kk^\prime}\frac{p(k|\bc)p(k^\prime|\bc)kk^\prime}{2\overline{k}(\bc)}~\Pic \log\Pik
\\
&&
+ \sum_k p(k|\bc)\log p(k)
\label{eq:OmegaPiPi}
\end{eqnarray}
with $\overline{k}(\bc)=\sum_k kp(k|\bc)$, $\Pi(\bc)=\{\Pi(k,k^\prime|\bc)\}$, and $p(\bc)=\{p(k|\bc)\}$.
The leading order $\frac{1}{2}\overline{k}(\bc)\log N$ in $\Omega(\ci|Q,p)$ reflects the property that the number of finitely connected graphs grows asymptotically with $N$ as $\exp[\sim\!\!N\log N]$. The next order is found to depend only on the macroscopic characterization
 $\{p(\bc),\Pi(\bc)\}$ of the {\em specific} graph $\bc$, and on the macroscopic characterization
 $\{p,\Pi\}$ of {\em typical} graphs from (\ref{eq:ensemble}), with $\Pi$ calculated for the kernel $Q$ via (\ref{eq:Pik}).

\subsection{Existence and uniqueness of tailored ensembles}

We will now prove that
for each degree distribution $p(k)$ and each relative degree correlation function $\Pi(k,k^\prime)$ there exist kernels $Q(k,k^\prime)$ such that their associated ensembles (\ref{eq:ensemble}) will for large $N$ be tailored to the production of random graphs with precisely these statistical features.
We identify these kernels and show that they all correspond in leading order in $N$ to the {\em same} random graph ensemble.
\begin{itemize}
\item
{\em Existence of a family of tailored kernels:}    \\[2mm]
For each non-negative function $\phi(k)$ such that $p(k)\phi(k)\Pi(k,k^\prime)\phi(k^\prime)p(k^\prime)$ is nonzero  for at least one combination $(k,k^\prime)$, the following kernel satisfies all conditions required to define a random graph ensemble of the family (\ref{eq:ensemble}) that generates graphs with degree distribution $p(k)$ and  relative degree correlation function $\Pi(k,k^\prime)$ as $N\to\infty$:
\begin{eqnarray}
\hspace*{-15mm}
Q(k,k^\prime)&=& \frac{\phi(k)\Pi(k,k^\prime)\phi(k^\prime)}{Z},~~~~~~~~Z=\sum_{kk^\prime}p(k)\phi(k)\Pi(k,k^\prime)\phi(k^\prime)p(k^\prime)
\label{eq:tailored_kernels}
\end{eqnarray}
$Q(k,k^\prime)$ is by construction non-negative, symmetric, and correctly normalized. Also we will always find $Z>0$ due to $\Pi(k,k^\prime)\geq 0$ in combination with our conditions on $\phi(k)$ and the normalization
(\ref{eq:Pinorm}). Recovering the correct degree distribution is built into the ensemble (\ref{eq:ensemble}) via the degree constraints.
To prove that equations (\ref{eq:Pik},\ref{eq:selfF}) are satisfied we define $F(k|Q)=\phi(k)/\sqrt{Z}$, and use the fact that by virtue of (\ref{eq:Pik}) the condition (\ref{eq:selfF}) reduces to (\ref{eq:selfPic}), and is therefore guaranteed to hold, provided $\Pi(k,k^\prime)$ indeed represents a relative degree correlation function.
\\[3mm]
What remains is to show that there exist functions $\phi(k)$ that meet the relevant conditions.
The simplest candidate is $\phi(k)=k/\bra k\ket$, for which we find $Z=1$ via (\ref{eq:Pinorm}) and which is easily confirmed to meet all criteria. It gives what we will call the {\em canonical kernel}:
\begin{eqnarray}
Q^\star(k,k^\prime)&=& \Pi(k,k^\prime)kk^\prime/\bra k\ket^2
\label{eq:Qcanonical}
\end{eqnarray}
\item {\em Completeness of the family of tailored kernels:}
\\[2mm]
The set of kernels defined by (\ref{eq:tailored_kernels}) is {\em complete}: if a kernel $Q(k,k^\prime)$ generates random graphs with
statistics $p(k)$ and $\Pi(k,k^\prime)$, then is must be of the form (\ref{eq:tailored_kernels}).
\\[2mm]
The proof is simple. If $Q(k,k^\prime)$ generates graphs with relative degree correlation function $\Pi(k,k^\prime)$,  according to (\ref{eq:Pik}) it must be of the form $Q(k,k^\prime) =F(k)\Pi(k,k^\prime) F(k^\prime)$ for some function $F(k)$. Since both $\Pi(k,k^\prime)$ and $Q(k,k^\prime)$ must be non-negative, the same must be true for $F(k)$. Hence $Q(k,k^\prime)$ is also of the form (\ref{eq:tailored_kernels}), with $\phi(k)=\sqrt{Z}F(k)$ and with the formula for $Z$ in (\ref{eq:tailored_kernels}) satisfied automatically due to $Q(k,k^\prime)$ having to be normalized.
\\[2mm]
A further corollary is that all kernels tailored to the generation of graphs with statistics $p(k)$ and $\Pi(k,k^\prime)$ are related to the canonical kernel (\ref{eq:Qcanonical}) via separable transformations,  with suitably normalized non-negative functions $G(k)$:
\begin{eqnarray}
Q(k,k^\prime)=G(k)Q^\star(k,k^\prime)G(k^\prime)
\label{eq:separable_trans}
\end{eqnarray}
\item {\em Asymptotic uniqueness of the canonical ensemble:}
\\[2mm]
The random graph ensembles of all kernels of the family (\ref{eq:tailored_kernels}), tailored to generating random graphs with statistical properties $p(k)$
and $\Pi(k,k^\prime)$, are asymptotically (i.e. for large enough $N$) identical: if all $\{k_i\}$ are drawn randomly from $p(k)$,
and $Q(k,k^\prime)$ belongs to the family (\ref{eq:tailored_kernels}) with canonical member $Q^\star(k,k^\prime)$ defined in (\ref{eq:Qcanonical}), then
\begin{eqnarray}
 [\Proba({\bf c}|p,Q)]^{1/N}= [\Proba({\bf c}|p,Q^\star)]^{1/N}e^{{\sl o}(1)}
\end{eqnarray}
This follows from (\ref{eq:asymptotic_ensemble}), which tells us that in the two leading orders in $N$ the probabilities of graphs generated from (\ref{eq:tailored_kernels}) depend on the kernel $Q(k,k^\prime)$ of the ensemble  only via its associated function $\Pi(k,k^\prime)$, so that
$N^{-1}\log \Proba(\ci|p,Q)-N^{-1}\log \Proba(\ci|p,Q^\star)={\sl o}(1)$.
\end{itemize}

The above results imply that we may regard the random graph ensemble (\ref{eq:ensemble}), equipped with the kernel (\ref{eq:Qcanonical}), as the natural ensemble for generating large random graphs with topologies controlled strictly by a prescribed degree distribution $p(k)$
and prescribed relative degree correlations $\Pi(k,k^\prime)$.
We will call $p(\ci|p,Q^\star)$, with $Q^\star(k,k^\prime)=\Pi(k,k^\prime)kk^\prime/\bra k\ket^2$,  the {\em canonical ensemble} for graphs with $p(k)$ and $\Pi(k,k^\prime)$. Note that for $\Pi(k,k^\prime)=1$ one has $Q^\star(k,k^\prime)=kk^\prime/\bra k\ket^2$, which is indeed equivalent to the trivial choice $Q(k,k^\prime)=1$ (as it is related to the latter by a separable transformation).

We can now also define what we mean by `null models'. Given the hypothesis that a network $\bc$ has no structure beyond that imposed by its degree statistics, the appropriate null model is a random graph generated by the canonical ensemble with degree distribution $p(k)=p(k|\bc)$ and relative degree correlations $\Pi(k,k^\prime)=1$ (giving the trivial kernel $Q(k,k^\prime)=1$; these are usually referred to as `simple graphs').
Similarly, given the  hypothesis that a network has no structure beyond that imposed by its degree statistics and its degree-degree correlations, the appropriate null model is a random graph generated by the canonical ensemble with degree distribution $p(k)=p(k|\bc)$ and
relative degree correlations $\Pi(k,k^\prime)=\Pi(k,k^\prime|\bc)$.
\vsp

Finally,
self-consistency demands that $p(k)$ and the canonical kernel (or a member of its equivalent family, related by separable transformations)
are also the most probable pair $\{p,Q\}$ in a Bayesian sense. The probability $\Proba(p,Q|\ci)$ that a pair $\{p,Q\}$
was the `generator' of $\bc$ via (\ref{eq:ensemble}) can be  expressed, via standard Bayesian relations, in terms of the probability
$\Proba(\ci|p,Q)$ of drawing $\bc$ at random from (\ref{eq:ensemble}):
\be
\Proba(p,Q|\ci)=\frac{\Proba(\ci,p,Q)}{\Proba(\ci)}=
\frac{\Proba(\ci|p,Q)\Proba(p,Q)}{\sum_{Q^\prime}\sum_{p^\prime} \Proba(\ci|p^\prime,Q^\prime,)\Proba(p^\prime,Q^\prime)}
\ee
The most probable pair $\{p,Q\}$ is the one that maximizes
$\log \Proba(p,Q|\bc)=\log \Proba(p,Q)+\log \Proba(\ci|p,Q)$ (modulo terms independent of $\{p,Q\}$),
so in the absence of any prior bias, i.e. if  $\Proba(p,Q)$ is independent of $\{p,Q\}$, it is the kernel that
maximizes $\Proba(\ci|p,Q)$. Since $\sum_{\bc}\Proba(\ci|p,Q)=1$ for any $\{p,Q\}$, finding the most probable $\{p,Q\}$ for a graph
$\ci$ boils down to finding the {\em smallest} ensemble of graphs compatible with the
structure of $\ci$.
Intuitively this makes sense: a more detailed characterization of the topology of an observed
graph allows for more information being carried over from the
graph to the ensemble, reducing the number of potential graphs allowed for by the ensemble.
The smaller the number of graphs in the ensemble, the more accurate will these graphs be when used as proxies for the observed one.

Maximizing $\Proba(\ci|p,Q)$ over $\{p,Q\}$ means minimizing $\Omega(\ci|p,Q)$ in (\ref{eq:Omega_def}), of which the leading orders in $N$ are given in
(\ref{eq:asymptotic_ensemble}).
Demonstrating Bayesian self-consistency of our canonical graph ensemble
for large $N$ hence boils down to proving that the maximum of  (\ref{eq:OmegaPiPi}) over  $\{p,\Pi\}$
(subject to the relevant constraints) is obtained for $\{p,\Pi\}=\{p(\bc),\Pi(\bc)\}$.
The constraints include the set (\ref{eq:selfPic}). There are clearly more,
e.g. $\Pi(k,k^\prime)\geq 0$ for all $(k,k^\prime)$,
however we show below
that maximizing (\ref{eq:OmegaPiPi}) over $\{p,\Pi\}$  subject only to (\ref{eq:selfPic}) and $\sum_k p(k)=1$
already generates the desired result: $\{p,\Pi\}=\{p(\bc),\Pi(\bc)\}$.
Extremizing\eq{eq:OmegaPiPi} with the Lagrange formalism, leads to the following
equations, which are to be solved in combination with (\ref{eq:selfPic}) and $\sum_k p(k)=1$:
\begin{eqnarray}
\hspace*{-20mm}
\forall(k,k^\prime):&~~\frac{\partial}{\partial\Pi(k,k^\prime)}\Omega[p(\bc),\Pi(\bc);p,\Pi]
&=\sum_{\ell\geq 0} \lambda(\ell)\frac{\partial}{\partial\Pi(k,k^\prime)}
\left(\frac{1}{\kav}\sum_{\ell^\prime} \ell^\prime p(\ell^\prime)\Pi(\ell,\ell^\prime)\!-\!1\right)
~
\label{eq:Lagrange1}
\\
\hspace*{-20mm}
\forall k:&~~\frac{\partial}{\partial p(k)}\Omega[p(\bc),\Pi(\bc);p,\Pi]
&=\sum_{\ell\geq 0} \lambda(\ell)\frac{\partial}{\partial p(k)}
\left(\frac{1}{\kav}\sum_{\ell^\prime} \ell^\prime p(\ell^\prime)\Pi(\ell,\ell^\prime)\!-\!1\right)
\nonumber
\\
\hspace*{-20mm}
&&\hspace*{20mm}
+\mu\frac{\partial}{\partial p(k)}\left(\sum_{k^\prime}p(k^\prime)\!-\!1\right)
\label{eq:Lagrange2}
\end{eqnarray}
in which $\{\lambda(\ell)\}$ and $\mu$ are Lagrange multipliers. Working out (\ref{eq:Lagrange1}) gives
\begin{eqnarray}
\forall(k,k^\prime):&~~~&
\frac{p(k)\Pi(k,k^\prime)p(k^\prime)}{\bra k\ket}=\frac{p(k|\bc)\Pi(k,k^\prime|\bc)p(k^\prime|\bc)}{\overline{k}(\bc)}
\frac{p(k)k}{2\lambda(k)}
\label{eq:selfcon1}
\end{eqnarray}
Since both $\Pik$ and $\Pic$ must satisfy the constraints (\ref{eq:selfPic}), with degree distributions $p(k)$ and $p(k|\bc)$, respectively,
it follows from (\ref{eq:selfcon1}) that
\begin{eqnarray}
\lambda(k)&=&\frac{1}{2}p(k)k
\label{eq:multipliers}
\end{eqnarray}
With this expression we eliminate $\lambda(k)$ from (\ref{eq:selfcon1}) to find
\begin{eqnarray}
\forall(k,k^\prime):&~~~&
\frac{p(k)\Pi(k,k^\prime)p(k^\prime)}{\bra k\ket}=\frac{p(k|\bc)\Pi(k,k^\prime|\bc)p(k^\prime|\bc)}{\overline{k}(\bc)}
\label{eq:selfcon2}
\end{eqnarray}
Next we work out (\ref{eq:Lagrange2}) and substitute (\ref{eq:multipliers}) into the result. This gives, using symmetry of $\Pi$:
\begin{eqnarray}
\forall k:~~~p(k|\bc)&=&
\mu p(k)+\frac{p(k)}{2\bra k\ket} \sum_{k^\prime} p(k^\prime)k^\prime \Pi(k^\prime,k)
= \Big(\mu+\frac{1}{2}\Big) p(k)
\end{eqnarray}
The normalization conditions $\sum_kp(k|\bc)=\sum_k p(k)=1$ then tell us that $\mu=\frac{1}{2}$, so $p(k)=p(k|\bc)$ for all $k$, and finally also (via
(\ref{eq:selfcon2})):
\begin{eqnarray}
\forall(k,k^\prime):&~~~&\Pik=\Pic
\label{eq:selfcon3}
\end{eqnarray}
Hence, the choice $\{p,\Pi\}=\{p(\bc),\Pi(\bc)\}$ indeed extremizes
the leading two orders in $N$ of (\ref{eq:asymptotic_ensemble}), subject to (\ref{eq:selfPic}) and to normalization of $p(k)$.
The above extremum must be a
maximum, since by making pathological choices for $\{p,\Pi\}$ (viz. choices inconsistent with the structure of $\bc$)
we can make $\Prob(\bc|p,Q)$ arbitrary small, and hence $\Omega[p(\bc),\Pi(\bc);p,\Pi]$ arbitrarily small.
Hence our canonical ensembles are indeed self-consistent in a Bayesian sense, as expected.

\subsection{The random graphs ensemble as a conditioned maximum entropy ensemble}

In this section we show that our canonical ensemble gives the maximum entropy within the subspace of graphs with prescribed degrees and upon
imposing as a constraint the average
values $\Pi(k,k')=\lav \Pi(k,k'|\bc)\rav$ of the relative degree correlations.
First we define our constraining observables, i.e. the degree sequence and
the re-scaled degree correlation:
\begin{eqnarray}
k_i(\bc)&=& k_i~~~~~~~~~~(\forall i)\\
q(k,k^\prime|\bc)&=& N^{-1}\sum_{ij}c_{ij}\dek\dekp~~~~~~(k,k^\prime >0)
\end{eqnarray}
Note that if $N^{-1}\sum_i \delta_{k,k_i(\bc)}=p(k)$ and $\kav=\sum_k p(k)$, then
\begin{eqnarray}
\Pi(k,k^\prime|\bc)&=&\frac{\bra k\ket}{{p(k)p(k^\prime)kk^\prime}}~q(k,k^\prime|\bc) ~~~~~{\rm for~}N\to\infty
\end{eqnarray}
We are interested in the maximum entropy random graph ensemble $p(\bc)$ (limited to symmetric graphs without self-interactions) such that
$q(k,k^\prime)=\sum_{\bc}p(\bc)q(k,k^\prime|\bc)$ for all $(k,k^\prime)$ and $k_i=k_i(\bc)$ for all $i$.
This is given by the ensemble $p(\bc)$ for which the Shannon entropy
\be
S[{\bf k},q]=\sum_{\bc}p(\bc|{\bf k},q)\log p(\bc|{\bf k},q)
\label{Shannon}
\ee
is maximal subject to our constraints.
Extremization of\eq{Shannon} with Lagrange multipliers gives, without
enforcing $p(\bc)\geq 0$ explicitly,
\begin{eqnarray}
\forall\bc:~~
\frac{\partial}{\partial p(\bc)}\left\{\sum_{\bc^\prime}p(\bc^\prime)\Big[\log p(\bc^\prime)+\Lambda_0
+\sum_{kk^\prime} \lambda(k,k^\prime) q(k,k^\prime|\bc^\prime)\Big]\right\}=0
\\
\forall\bc:~~
\log p(\bc)+\Lambda_0
+\sum_{kk^\prime} \lambda(k,k^\prime) q(k,k^\prime|\bc)+1=0
\\
\forall\bc:~~
p(\bc)=\frac{1}{\cal Z}~e^{
-\sum_{kk^\prime} \lambda(k,k^\prime) q(k,k^\prime|\bc)}
\\
\forall\bc:~~
p(\bc)=\frac{1}{\cal Z}~e^{
-N^{-1}\sum_{ij}c_{ij}\lambda(k_i(\bc),k_j(\bc))}
\label{exponential}
\end{eqnarray}
with $\Partition$ such that $\sum_{\bc}p(\bc)=1$.
As expected for an ensemble of random graphs
with maximum entropy, where a set of averages
of obervables are constrained to assume prescribed values, the result of the
extremization gives an exponential family,
where the parameters $\{\lambda(k_i,k_j)\}$
are to be calculated from the equations for the constraints.
What is left is to show that the exponential family can be reduced to the
micro-canonical ensemble \eq{eq:micro_ensemble}, where degrees are prescribed,
by a simple redefinition of the
Lagrange multipliers.
Let us first rewrite\eq{exponential}
\bea
\forall\bc:~~
p(\bc)=\frac{1}{\cal Z}
\Big(\prod_{i<j}e^{-N^{-1}c_{ij}[\lambda(k_i,k_j)+\lambda(k_j,k_i)]}\Big) \Big(\prod_i \delta_{k_i,k_i(\bc)}\Big)
\\
\forall\bc:~~
p(\bc)=\frac{1}{\cal Z}
\Big(\prod_{i<j}\Big[e^{-N^{-1}[\lambda(k_i,k_j)+\lambda(k_j,k_i)]}\delta_{c_{ij},1}
+\delta_{c_{ij},0}
\Big]\Big) \Big(\prod_i \delta_{k_i,k_i(\bc)}\Big)
\eea
We can then redefine our Langrange multipliers in terms of the function
$Q(k,k^\prime)$ via
\begin{eqnarray*}
\frac{\bra k\ket}{N} Q(k,k^\prime)&=&\frac{e^{-N^{-1}[\lambda(k_i,k_j)+\lambda(k_j,k_i)]}}{1+e^{-N^{-1}[\lambda(k_i,k_j)+\lambda(k_j,k_i)]}}
\end{eqnarray*}
This results in
\begin{eqnarray}
p(\bc)&=&\frac{1}{\cal Z}
\prod_{i<j}\Big(1-\frac{\bra k\ket Q(k_i,k_j)}{N}\Big)^{-1}
\nonumber\\
&& \times \prod_{i<j}\Big[\frac{\bra k\ket}{N}Q(k_i,k_j)\delta_{c_{ij},1}
+\Big(1-\frac{\bra k\ket}{N}Q(k_i,k_j)\Big)\delta_{c_{ij},0}
\Big].\prod_i \delta_{k_i,k_i(\bc)}
\label{eq:pc}
\end{eqnarray}
The first product in\eq{eq:pc} only depends on the constrained degrees $\{k_i\}$
(in fact, to leading order this dependence is only via their average $\kav$, since
$\prod_{i<j}(1-\bra k\ket Q(k_i,k_j)/N)^{-1}=e^{N\kav/2+\order{(1)}}$), so
it drops out of the measure,
and hence \eq{eq:pc} can be rewritten as
\begin{eqnarray}
\hspace*{-15mm}
p(\bc)&=& \frac{1}{\Partition(\bk,Q)}\prod_{i<j}\left[
\frac{\bra k\ket}{N}\Qij\de+\left(1\!-\!\frac{\bra k\ket}{N}\Qij\right)\!\deO
\right]\prod_i\deki
\\
\hspace*{-15mm}
\Partition(\bk,Q)&=&\sum_{\bc} \prod_{i<j}\left[
\frac{\bra k\ket}{N}\Qij\de+\left(1\!-\!\frac{\bra k\ket}{N}\Qij\right)\!\deO
\right]\prod_i\deki
\label{zeta}
\end{eqnarray}
which indeed reduces to\eq{eq:micro_ensemble}, as claimed.


\subsection{Shannon entropy}

The (rescaled) Shannon entropy of the canonical ensemble
$\Proba(\ci|Q^\star,p)$, as defined by $Q^\star(k,k^\prime)=\Pi(k,k^\prime)kk^\prime/\bra k\ket^2$ in combination with (\ref{eq:ensemble}), is an important quantity as it allows us to define and calculate the effective number of graphs ${\cal N}[p,\Pi]$ in the ensemble:
\begin{eqnarray}
S[p,\Pi]&=&-\frac{1}{N}\sum_{\ci}\Proba(\ci|p,Q^\star)~ \log \Proba(\ci|p,Q^\star)
\label{eq:shannon}
\\
{\cal N}[p,\Pi]&=& e^{NS[p,\Pi]}
\label{eq:nr_of_graphs}
\end{eqnarray}
In (\ref{eq:shannon}) one defines as always $0\log 0=\lim_{\epsilon\downarrow 0}\epsilon\log\epsilon=0$.
For large $N$ we can use our earlier results (\ref{eq:Omega_def},\ref{eq:asymptotic_ensemble},\ref{eq:OmegaPiPi})
to find the leading orders of the entropy, since
\begin{eqnarray}
S[p,\Pi]&=&\sum_{\ci}\Proba(\ci|p,Q^\star)
\Omega(\ci|p,Q^\star)
\nonumber
\\
&=&
\frac{1}{2}\bra k\ket\Big[\log[ N\bra k\ket]\! -\!1\Big] -\sum_k p(k)\log k!
\nonumber
\\
&&
-\sum_{\ci}\Proba(\ci|p,Q^\star)\Omega[\Pi(\bc),p(\bc);\Pi,p]+{\sl o}(1)
\nonumber
\\
&=&
\frac{1}{2}\bra k\ket\Big[\log[ N\bra k\ket]\! -\!1\Big] -\sum_k p(k)\log k!- \sum_k p(k)\log p(k)
\nonumber
\\
&&
-\sum_{kk^\prime}\frac{p(k)p(k^\prime)kk^\prime}{2\bra k\ket}\Pik \log\Pik
+{\sl o}(1)
\nonumber
\\
&=&
\frac{1}{2}\bra k\ket\Big[\log[ N/\bra k\ket]\! +\!1\Big] - \sum_k p(k)\log[ p(k)/\pi(k)]
\nonumber
\\
&&
-\frac{1}{2\bra k\ket}\sum_{kk^\prime}p(k)p(k^\prime)kk^\prime\Pik \log\Pik
+{\sl o}(1)
\label{eq:Sintermediate1}
\end{eqnarray}
where $\pi(k)$ denotes the Poissonian degree distribution with average degree $\bra k\ket$, viz. $\pi(k)=e^{-\bra k\ket}\bra k\ket^k/k!$.
To prove various properties of the above expression for the entropy it will be convenient to introduce a new (symmetric) quantity $W(k,k^\prime)$, defined as the probability that a randomly drawn link in a graph that has $\Pi(k,k^\prime|\bc)=\Pi(k,k^\prime)$ connects two nodes with degrees $k$ and $k^\prime$.  It can be shown to be related to
$\Pi(k,k^\prime)$ via
\begin{eqnarray}
W(k,k^\prime)&=& p(k)p(k^\prime)kk^\prime\Pi(k,k^\prime) /\bra k\ket^2
\label{eq:defW}
\end{eqnarray}
Irrespective of its exact meaning, the crucial mathematical advantage here of working with $W(k,k^\prime|p,\Pi)$ is that it represents a probability distribution: $W(k,k^\prime)\geq 0$ and $\sum_{kk^\prime}W(k,k^\prime)=1$
(normalization follows from (\ref{eq:selfPic})). One also verifies explicitly that $W(k)=\sum_{k^\prime} W(k,k^\prime)=p(k)k/\bra k\ket$ for all $k$. If we use (\ref{eq:defW}) to eliminate $\Pi(k,k^\prime)$ from (\ref{eq:Sintermediate1}) in favour of
$W(k,k^\prime)$ we get
\begin{eqnarray}
S[p,\Pi]
&=&
\frac{1}{2}\bra k\ket\Big[\log[ N/\bra k\ket]\! +\!1\Big] - \sum_k p(k)\log[ p(k)/\pi(k)]
\nonumber
\\
\hspace*{-15mm}&&
-\frac{1}{2}\bra k\ket\sum_{kk^\prime} W(k,k^\prime)\log
\Big[W(k,k^\prime)/W(k)W(k^\prime)\Big]
+{\sl o}(1)
\label{eq:Sintermediate2}
\end{eqnarray}
The term in (\ref{eq:Sintermediate2}) with $W(k,k^\prime)$ is seen to be proportional to minus the mutual information between two connected sites, and is therefore non-positive, vanishing if and  only if $\Pi(k,k^\prime)=1$ for all $(k,k^\prime)$.
Furthermore, the term in (\ref{eq:Sintermediate2}) involving $\pi(k)$ is minus a KL-divergence, and therefore also non-positive, vanishing if and only if $p(k)=\pi(k)$ for all $k$.
Our result (\ref{eq:Sintermediate2}) therefore has a clear and elegant interpretation:
\begin{itemize}
\item
For the simplest graphs of the Erd\"{o}s-R\'{e}nyi type, where only the average degree $\bra k\ket$ is imposed,
one has $p(k)=\pi(k)$ for all $k$ and $\Pi(k,k^\prime)=1$ for all $(k,k^\prime)$. This gives $W(k,k^\prime)=W(k)W(k^\prime)$, and the entropy
takes its maximal value:
\begin{eqnarray}
 S[p,\Pi]&=&\frac{1}{2}\bra k\ket\Big[\log[ N/\bra k\ket]\! +\!1\Big]
\label{ErdosRenyi}
\end{eqnarray}
\item
For graphs where the degree distribution $p(k)$ is imposed, but without further structure (i.e. still
$\Pi(k,k^\prime)=1$ for all $(k,k^\prime)$),
the entropy decreases by an amount $\sum_k p(k)\log[ p(k)/\pi(k)]$ which is the KL-distance between the imposed $p(k)$ and the Poissonian degree
distribution with the same average connectivity:
\begin{eqnarray}
S[p,\Pi]&=&\frac{1}{2}\bra k\ket\Big[\log[ N/\bra k\ket]\! +\!1\Big]
- \sum_k p(k)\log[ p(k)/\pi(k)]
\end{eqnarray}
\item
For the more sophisticated graphs where both a degree distribution $p(k)$ and nontrivial
degree correlations defined via $\Pi(k,k^\prime)$ are imposed, one no longer has $W(k,k^\prime)=W(k)W(k^\prime)$ and
the entropy decreases further
by an amount $\frac{1}{2}\bra k\ket\sum_{kk^\prime} W(k,k^\prime)\log
\big[W(k,k^\prime)/W(k)W(k^\prime)\big]$, which is proportional to the mutual information regarding degrees of connected nodes:
\begin{eqnarray}
S[p,\Pi]&=&\frac{1}{2}\bra k\ket\Big[\log[ N/\bra k\ket]\! +\!1\Big]
- \sum_k p(k)\log[ p(k)/\pi(k)]
\nonumber
\\
&&
-\frac{1}{2}\bra k\ket\sum_{kk^\prime} W(k,k^\prime)\log
\Big[W(k,k^\prime)/W(k)W(k^\prime)\Big]
\label{totalEntropy}
\end{eqnarray}
\end{itemize}

\section{Quantitative tools for networks}

The availability for any given/observed
network $\bc$ of a well-defined canonical random graph ensemble, that produces random graphs with microscopic topologies
controlled solely by the observed degree statistics and degree correlations of the given $\bc$, allows us to develop practical
quantitative tools with which to analyze and compare (structure in) real networks. Here we focus on three such tools.

\subsection{Quantifying structural network complexity}

\begin{figure}[t]
\vspace*{5mm}
\center{\includegraphics[width=270\unitlength]{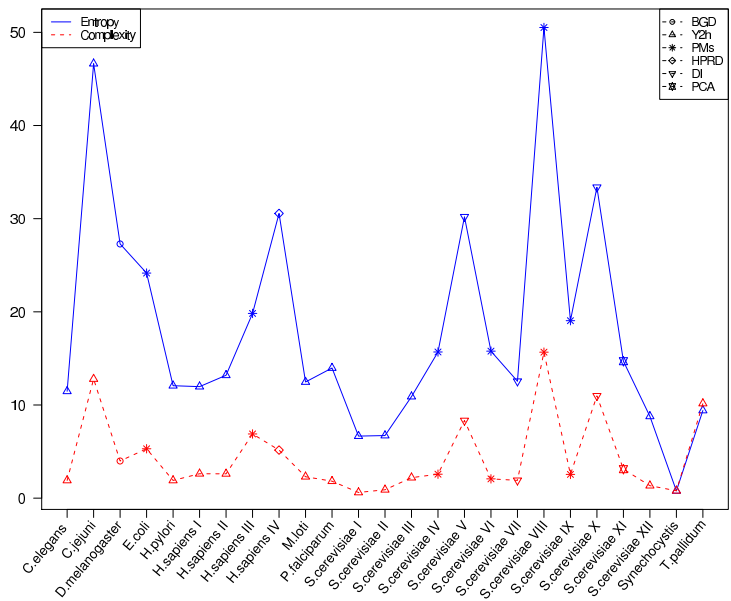}}
\vspace*{15mm}
\caption{Shannon entropy per node $S[p,\Pi]$ (markers connected by solid lines) and complexity ${\cal C}[p,\Pi]$
(markers connected by dashed lines)
of the canonical ensembles tailored to the
production of random graphs with microscopic topologies controlled solely
by the degree sequence and degree correlation of experimentally determined PPINs. 
The methods/sources for the experimental data sets are the following:
BGD, BioGrid database; Y2h, yeast two-hybrid screen; PMs, purification mass spectrometry; HPRD, human protein reference database; DI, data integration (database with combined experimental data); PCA, protein fragment complementation assay. The studied 
organisms are listed in alphabetical order on the x-axis. Data sets properties and references are summarised in Table~\ref{tab:properties}.}
\label{fig:Entropy}
\end{figure}

The natural definition of the complexity of a given network $\bc$ is based on the number ${\cal N}[p,\Pi]$ of graphs in its canonical ensemble $\{p,\Pi\}$, and hence on the entropy per node $S[p,\Pi]$ given in (\ref{eq:Sintermediate2}).
It makes sense to write this entropy for large $N$ as $S[p,\Pi]
=S_0 - {\cal C}[p,\Pi]+{\sl o}(1)$,  with a first (positive) contribution $S_0=\frac{1}{2}\bra k\ket\big[\log[ N/\bra k\ket\! +\!1\big]$  that originates simply from counting the total number of bonds and would also be found for
structureless Erd\"{o}s-R\'{e}nyi graphs (where only the average degree is prescribed), minus a second term ${\cal C}[p,\Pi]$
which acts to {\em reduce} the entropy as soon as there is structure in the graph beyond a prescribed average degree.
This latter quantity ${\cal C}[p,\Pi]$ can be identified as the complexity  of graphs in the canonical ensemble associated with $\bc$, and hence as the complexity of $\bc$:
\begin{eqnarray}
\hspace*{-10mm}
{\cal C}[p,\Pi]&=&
\sum_k p(k)\log[ p(k)/\pi(k)]
+\frac{1}{2\bra k\ket}\sum_{kk^\prime} p(k)p(k^\prime)kk^\prime\Pi(k,k^\prime) \log
\Pi(k,k^\prime)
\label{Complexity}
\end{eqnarray}
where $\pi(k)$ is the Poissonian distribution with average degree $\bra k\ket$:
\begin{eqnarray}
\pi(k)=e^{-\bra k\ket}\bra k\ket^k/k!
\end{eqnarray}
The larger ${\cal C}[p,\Pi]$, the more `rare' or `special' are graphs
with characteristics $\{p,\Pi\}$. For every $N$, the complexity is
bounded from above by\eq{ErdosRenyi}; at this value
the network undergoes an entropy `crisis', as \eq{totalEntropy}
vanishes and the degree distribution ceases to be graphical, \ie~ 
no network can be found with this degree distribution
(see \cite{ErdGal60} for the notion of graphicality).
Note, however, that
our results were obtained in the limit $\kav \ll N$; they
no longer apply for degree distributions with an average
connectivity of the order of the system size. For
e.g. fully connected graphs, where the complexity is maximal, the entropy
should vanish, whereas
\eq{totalEntropy} indeed yields an incorrect $\order{(N)}$ result.
As an illustration one may check how close to the entropy crisis are  PPIN of different species
(PPIN typically meet the requirements  $\bra k\ket \ll N$ for our theory to apply).
For this purpose we
have computed the \eq{totalEntropy} for protein interaction networks of
different species  and show the results in Fig.~\ref{fig:Entropy}.
A more systematic and extensive application of our tools to PPIN will be published in \cite{Fernandes}.
\begin{table}[H]
\centering
\begin{tabular}{lrrrrrrl}
\\
Species & $k_{\mbox{max}}$ & Method & Reference
\\
\hline
C.elegans & 99 & Y2h & \cite{Simonis08}\\
C.jejuni & 207 & Y2h & \cite{Parrish07}\\
D.melanogaster & 176 & BGD & \cite{Stark06}\\
E.coli & 641 & PMs & \cite{Arifuzzaman06}\\
H.pylori & 55 & Y2h & \cite{Rain01}\\
H.sapiens I & 125 & Y2h & \cite{Rual05}\\
H.sapiens II & 95 & Y2h & \cite{Stelzl05}\\
H.sapiens III & 314 & PMs & \cite{Ewing07}\\
H.sapiens IV & 247 & HPRD & \cite{Prasad09}\\
M.loti & 401 & Y2h & \cite{Shimoda08}\\
P.falciparum & 51 & Y2h & \cite{Lacount05}\\
S.cerevisiae I & 24 & Y2h & \cite{Uetz00}\\
S.cerevisiae II & 55 & Y2h & \cite{Itocore01}\\
S.cerevisiae III & 279 & Y2h & \cite{Itocore01}\\
S.cerevisiae IV & 62 & PMs & \cite{Ho02}\\
S.cerevisiae V & 118 & DI & \cite{VonMering02}\\
S.cerevisiae VI & 53 & PMs & \cite{Gavin02}\\
S.cerevisiae VII & 32 & DI & \cite{Han04}\\
S.cerevisiae VIII & 955 & PMs & \cite{Gavin06}\\
S.cerevisiae IX & 141 & PMs & \cite{Krogan06}\\
S.cerevisiae X & 127 & DI & \cite{Collins07}\\
S.cerevisiae XI & 58 & PCA & \cite{Tarassov08}\\
S.cerevisiae XII & 86 & Y2h-PCA & \cite{Yu08}\\
Synechocystis & 51 & Y2h & \cite{Sato07}\\
T.pallidum & 285 & Y2h & \cite{Titz08}\\
\hline
\end{tabular}
\caption{Maximum degree $k_{\mbox{max}}$, detection method/source and reference for the biological network data sets. 
The detection methods/sources are abbreviated as in Fig.~\ref{fig:Entropy}
}
\label{tab:properties}
\end{table}

\subsection{Quantifying structural distance between networks}

In the same spirit
we can now also use our tailored graph ensembles to define an information-theoretic distance $D_{AB}$ between any two networks $\bc_A$ and $\bc_B$, based solely on the macroscopic structure statistics as captured
by their associated (observed) structure function pairs $\{p_A,\Pi_A\}$ and   $\{p_B,\Pi_B\}$.
The natural definition would be in terms of the Jeffreys divergence (i.e. the symmetrized KL-distance) between the two associated canonical ensembles, which is non-negative and equals zero if and only if $\{p_A,\Pi_A\}=\{p_B,\Pi_B\}$, i.e. if the graphs $\bc_A$ and $\bc_B$ belong to the same canonical ensemble:
\begin{eqnarray}
D_{AB}&=& \frac{1}{2N}\sum_{\bc}\Proba(\bc|p_A,Q_A)\log \Big[\frac{\Proba(\bc|p_A,Q_A)}{\Proba(\bc|p_B,Q_B)}\Big]
\nonumber
\\
&&
+\frac{1}{2N}\sum_{\bc}\Proba(\bc|p_B,Q_B)\log \Big[\frac{\Proba(\bc|p_B,Q_B)}{\Proba(\bc|p_A,Q_A)}\Big]
\end{eqnarray}
Working out this formula, using (\ref{eq:asymptotic_ensemble}) and (\ref{eq:Sintermediate2}), gives
for large $N$:
\begin{eqnarray}
\hspace*{-15mm}
D_{AB}
&=&
 \frac{1}{2}\sum_{\bc}\Proba(\bc|p_A,Q_A)\Omega(\bc|p_B,Q_B)
-\frac{1}{2}S[p_A,\Pi_A]
\nonumber
\\
\hspace*{-20mm}
&&
 +\frac{1}{2}\sum_{\bc}\Proba(\bc|p_B,Q_B)\Omega(\bc|p_A,Q_A)
-\frac{1}{2}S[p_B,\Pi_B]
\nonumber
\\
\hspace*{-20mm}
&=&
 \frac{1}{2}\sum_k p_A(k)\log\Big[ \frac{p_A(k)}{p_B(k)}\Big]
 +\sum_{kk^\prime}\frac{p_A(k)p_A(k^\prime)kk^\prime}{4\bra k\ket_A}\Pi_A(k,k^\prime)\log\Big[\frac{\Pi_A(k,k^\prime)}{\Pi_B(k,k^\prime)}\Big]
 \label{eq:distance_formula}
 \\
 \hspace*{-20mm}
 &&
 + \frac{1}{2}\sum_k p_B(k)\log\Big[ \frac{p_B(k)}{p_A(k)}\Big]
+\sum_{kk^\prime}\frac{p_B(k)p_B(k^\prime)kk^\prime}{4\bra k\ket_B}\Pi_B(k,k^\prime)\log\Big[\frac{\Pi_B(k,k^\prime)}{\Pi_A(k,k^\prime)}\Big]
+{\sl o}(1)
\nonumber
\end{eqnarray}
This quantity is used in \cite{Fernandes} for comparing and clustering PPIN data sets, even if these differ in size,
solely on the basis of their degree sequence and degree correlations. The combination of its information-theoretic origin and explicit nature
(so that it involves almost no computational cost) makes (\ref{eq:distance_formula}) an efficient practical tool in bio-informatics.

\subsection{Numerical generation of canonical `null models'}

We have shown that for any given network $\ci$ it is possible to define
a tailored ensemble of graphs, that share with $\ci$ those structural aspects that follow directly from its degree distribution and degree correlations, and used it to define and calculate complexities and structural distances.  Our next aim is to use  the ensemble for
{\em generating} random graphs with structure $\{p,\Pi\}$ identical to that of a given network.
The problems associated with generating complex random graphs with controlled properties
are well known \cite{rao,Gka,Viger,chen,catanzaro,serrano,foster,verhelst1}. In \cite{CooDemAnn09}
a general method was proposed for generating random graphs with
built-in constraints and specific statistic weights, such as described by the
invariant measure \eq{eq:ensemble}, in the form of a Monte-Carlo process that is guaranteed to evolve
from any initial graph $\ci_0$ that meets the relevant constraints
towards the prescribed invariant measure\eq{eq:ensemble}.
The initial graph $\ci_0$ can be constructed by hand, for any choice of
$p(k)$, such that for sufficiently large $N$ it will have the
required degree statistics (see~\eg~\cite{NewStrWat01}).
With the general and exact algorithm \cite{CooDemAnn09} we can generate graphs according to the measure
\eq{eq:ensemble}, with the kernel $Q(k,k^\prime)$ of \eq{eq:Qcanonical}
chosen such as to impose any desired degree correlations $\Pi(k,k^\prime)$.
These graphs can then serve as `null models',
allowing us, for instance, to determine to what extent specific small motifs in biological networks (such as short loops)
can be regarded as mere consequences of the overall structure dictated by their degree statistics and degree correlations,
or whether they reflect deeper biological principles. See \cite{Fernandes} for the results of such tests.

\unitlength=0.25mm
\begin{figure}[t]
\hspace*{20mm}
\begin{picture}(200,460)
\put(110,450){\bf (a)}
\put(0,230){\includegraphics[width=250\unitlength]{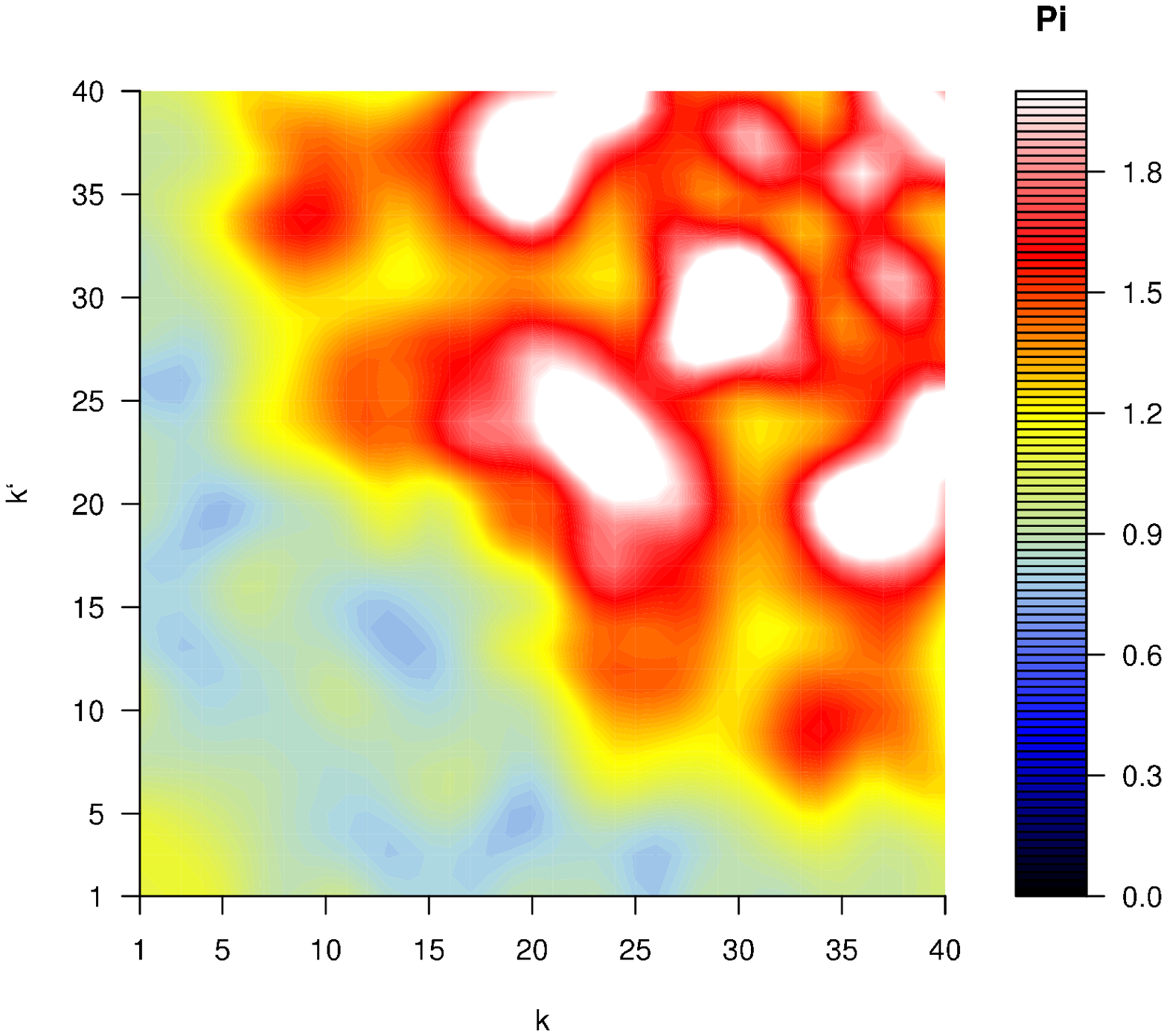}}
\put(400,450){\bf (b)}
\put(290,230){\includegraphics[width=250\unitlength]{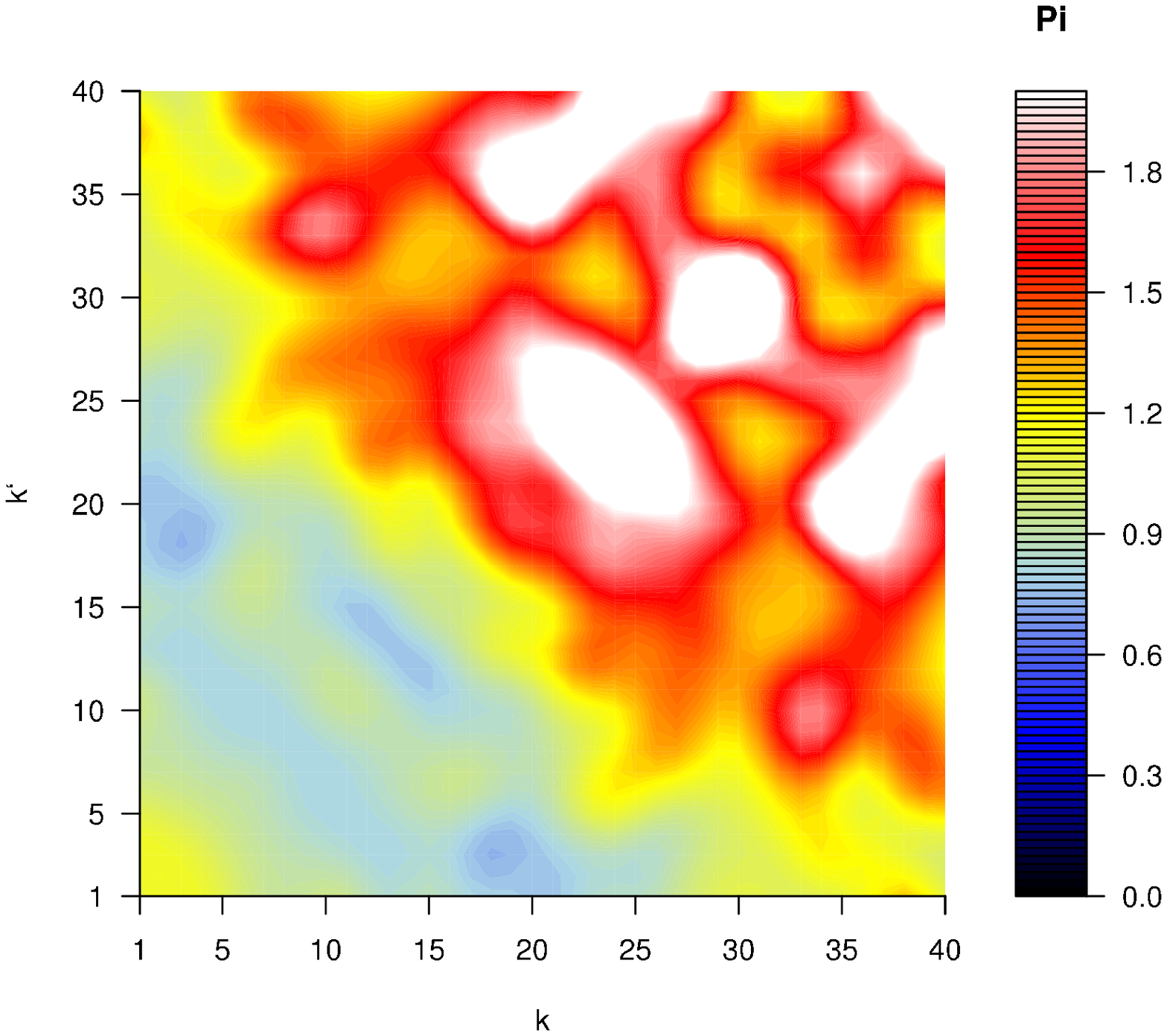}}
\put(260,220){\bf (c)}
\put(150,0){\includegraphics[width=250\unitlength]{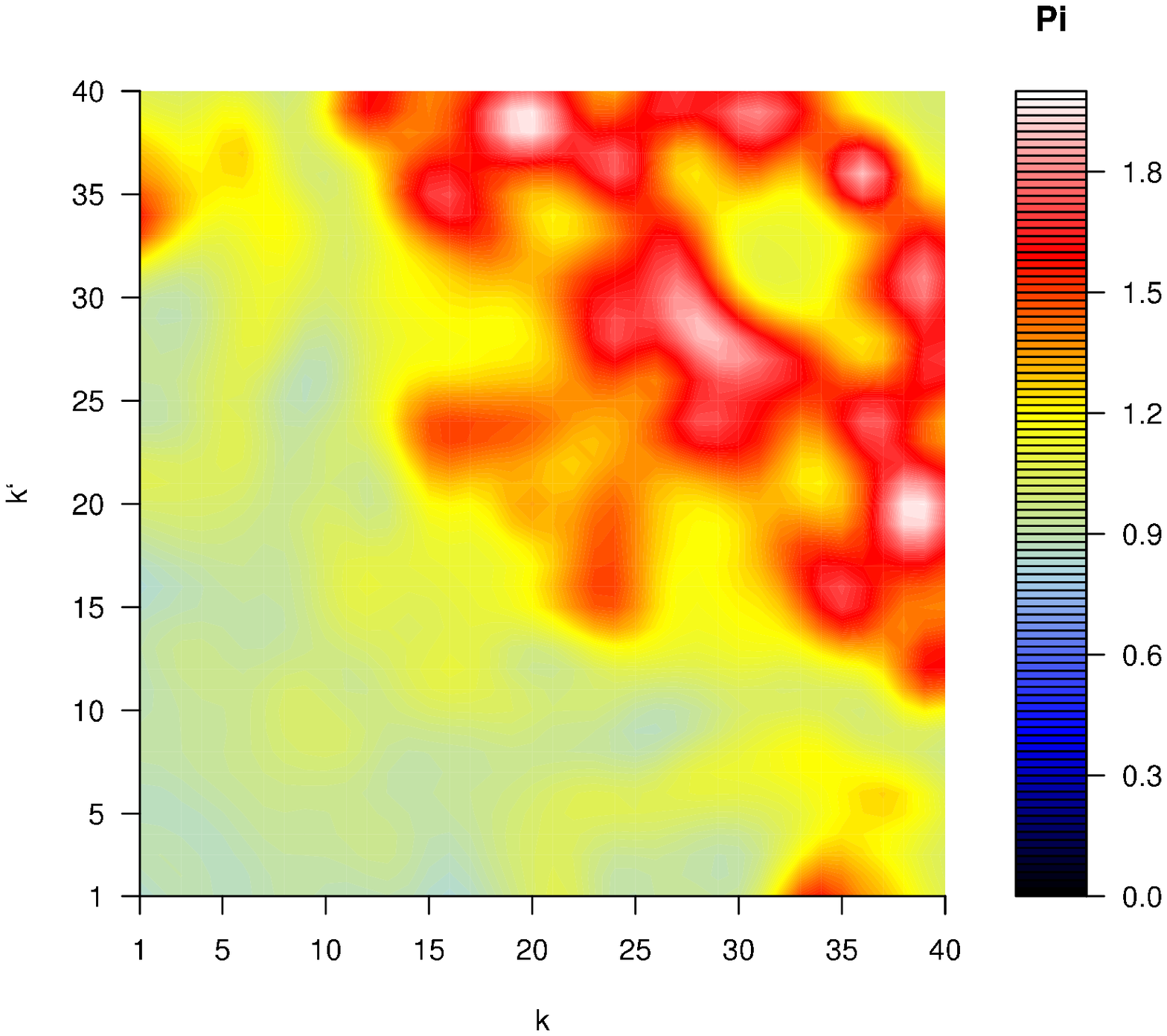}}
\end{picture}
\caption{Results of Markov chain graph dynamics proposed in \cite{CooDemAnn09} tailored to generating equilibrium
random graph ensembles with specific
degree sequences {\it and} specific degree correlations. (a): Colour plot of the
relative degree correlations $\Pi(k,k^\prime|\bc)$ as measured in the {\it Escherichia coli}
PPIN (here $N=2457$ and $\kav=7.05$).
(b): colour plot of $\Pi(k,k^\prime|\bc^\prime)$ in
the synthetic graph $\bc^\prime$ generated with Markov chain dynamics targeting the
measured degree correlation if the PPIN, after $67,147$ accepted moves.
(c): colour plot of $\Pi(k,k^\prime|\bc^\prime)$ in
the final graph generated with dynamics  targeting
$\Pi(k,k^\prime)=1~\forall~(k,k^\prime)$, after 1,968,000 accepted moves.}
\label{fig:Piecoli}
\end{figure}

Here we generate, as an illustration, a synthetic network which is to have the same
degree sequence and the same degree correlations as the protein interaction
network of {\it Escherichia coli}, as given in~\cite{Arifuzzaman06}, (i.e. we produce a member of the tailored graph ensemble of this particular PPIN),
where $N=2457$ and $\kav=7.05$.
The degree correlations of the resulting graph
after $67,\!147$
accepted moves of the Markov chain algorithm of \cite{CooDemAnn09} are shown in figure~\ref{fig:Piecoli}(b), and are seen to be in very good agreement
with the degree correlations of the PPIN that are being targeted, displayed in figure~\ref{fig:Piecoli}(a) (note that there is no need to compare degree distributions, since all degrees are guaranteed to be conserved by the graph dynamics \cite{CooDemAnn09}).
To rule out the possibility that the observed similarity in degree correlations between the synthetic graph and the original PPIN
could have arisen from poor sampling of the microscopic configurations (and just
reflect direct similarities in the connectivity matrices), we also calculated the Hamming
distance between the connectivity matrices $\bc$ and $\bc^\prime$ of the
original PPIN and the synthetic graph,
\be
\rho(\bc,\bc^{\prime})=\frac{1}{2 N \kav}\sum_{ij}|c_{ij}-c'_{ij}|,
\label{eq:Hamming}
\ee
(the prefactor is chosen such that when the
two matrices differ in all the $2N\kav$ entries which
could be different, then $\rho(\bc,\bc^{\prime})=1$). The Hamming distance vanishes if the two matrices are identical.
In the present case we find $\rho=0.90$, which implies that
although our two graphs have similar macroscopic structure,
their microscopic realizations are indeed very different.

For comparison, we also show in figure~\ref{fig:Piecoli}(c) the degree correlations of a
synthetic graph obtained via the Markov chain dynamics of \cite{CooDemAnn09}, starting from the same initial graph,
but now targeting degree correlations described by $\Pi(k,k^\prime)=1$ for all $(k,k^\prime)$.
All residual deviations in the bottom plot of figure \ref{fig:Piecoli} from the objective $\Pi(k,k')=1~\forall (k,k^\prime)$ are due to finite size effects.
Again we
also calculate the Hamming distance between the original
and the synthetic matrix, giving $\rho=0.93$. This value is similar to
the one found previously, but now the macroscopic structure of the
synthetic graph in terms of the degree correlations is considerably different from the underlying PPIN.

It has been noted by several authors that most PPINs are disassortative,
i.e. nodes with high degrees tend to connect with nodes with low
degrees~\cite{Newman02}. Measures of degree assortativity have been proposed in \cite{PasVazVes01,Newman02,Newman03}.
A conventional measure of assortativity is the correlation coefficient $(\bra kk^\prime\ket-\bra k\ket\bra k^\prime\ket)/(\bra k^2\ket-\bra k\ket^2)$, calculated over the joint distribution $W(k,k^\prime)$ in (\ref{eq:defW}).
Degree-assortativity has been shown to have important consequences on both
the topology of a network and the process which it supports. In particular,
it was shown that assortative networks are more resistant to
random attacks, i.e. random vertex removal,
whereas disassortative networks are less
resistant~\cite{Newman02}.
It may be useful from a practical point of view to generate networks
with a prescribed assortative character. This can again be achieved by using
the measure\eq{eq:ensemble}, where the kernels
$Q(k,k^\prime)$ are now chosen to produce assortative or disassortative graphs.
In~\cite{CooDemAnn09} it was shown that the kernel
\be
Q(k,k')=\frac{|k-k^\prime|^2}{2(\langle k^2\rangle-\kav^2)}
\label{disass_example}
\ee
tailors the ensemble\eq{eq:ensemble} to the production, for large $N$,
of graphs with degree correlations
\begin{eqnarray}
\Pi(k,k^\prime)&=& \frac{\bra k\ket (k-k^\prime)^2}{[\alpha_3-2\alpha_2 k+\alpha_1 k^2][\alpha_3-2\alpha_2 k^\prime+\alpha_1 k^{\prime 2}]}
\end{eqnarray}
where the three coefficients $\alpha_\ell$ are to be solved numerically from
\begin{eqnarray}
\alpha_\ell&=&\sum_k  \frac{k^{\ell} p(k)}{\alpha_3-2\alpha_2 k+\alpha_1 k^2}
\label{eq:alphas}
\end{eqnarray}
This degree correlation has a disassortative character.
In fact, any kernel of the form
\begin{eqnarray}
Q(k,k^\prime)&=&C^{-1}{|k-k'|^n},~ n=1,2,\ldots
\label{eq:disassortative}
\end{eqnarray}
with $C$ such that $\sum_{k,k^\prime}p(k)p(k^\prime)Q(k,k^\prime)=1$
will tailor the ensemble\eq{eq:ensemble}, for large $N$,
to the production
of graphs with increasingly negative assortative coefficients
as $n$ increases.
A prototype of an assortative kernel would be
\begin{eqnarray}
Q(k,k^\prime)= \frac{1}{C}\frac{1}{1+|k-k^\prime|^n}
\label{eq:assortative}
\end{eqnarray}
where $C=\sum_{k,k^\prime}p(k)p(k^\prime)[1+|k-k^\prime|^n]$.
For sufficiently large $N$, the predicted values for $\Pi(k,k^\prime)$ follow
from\eq{eq:Pik},
where $F(k)$ is to be solved numerically from (\ref{eq:selfF}).
As an example we generated
two synthetic graphs, both with the same
degree sequence as the PPIN of {\it Homo sapiens} 
(the experimental data used was taken from the human protein reference database (HPRD),~\cite{Prasad09}). In the first graph we enforced
an assortative connectivity using (\ref{eq:assortative}) with $n=1$, and in the second one a disassortative connectivity
using (\ref{eq:disassortative}) with $n=1$. Both graphs
were generated with the algorithm of \cite{CooDemAnn09}, starting
from the actual {\it Homo sapiens} PPIN. In Fig.~\ref{fig:Pisapiens}(a) we show the colour plot of
the relative degree correlations $\Pi(k,k^\prime)$ as measured in the {\it Homo sapiens} PPIN,
and in Fig.~\ref{fig:Pisapiens}(c) and~\ref{fig:Pisapiens}(e) we show the same quantity in the two synthethic graphs generated. 
For comparison we also show (Fig.~\ref{fig:Pisapiens}(b) and~\ref{fig:Pisapiens}(d), respectively) the
functional $\Pi(k,k^\prime)$ that are being targeted, via the kernels in (\ref{eq:assortative}) and (\ref{eq:disassortative}).

\unitlength=0.25mm
\begin{figure}[t]
\hspace*{20mm}
\begin{picture}(200,710)
\put(270,680){\bf (a)}
\put(150,460){\includegraphics[width=250\unitlength]{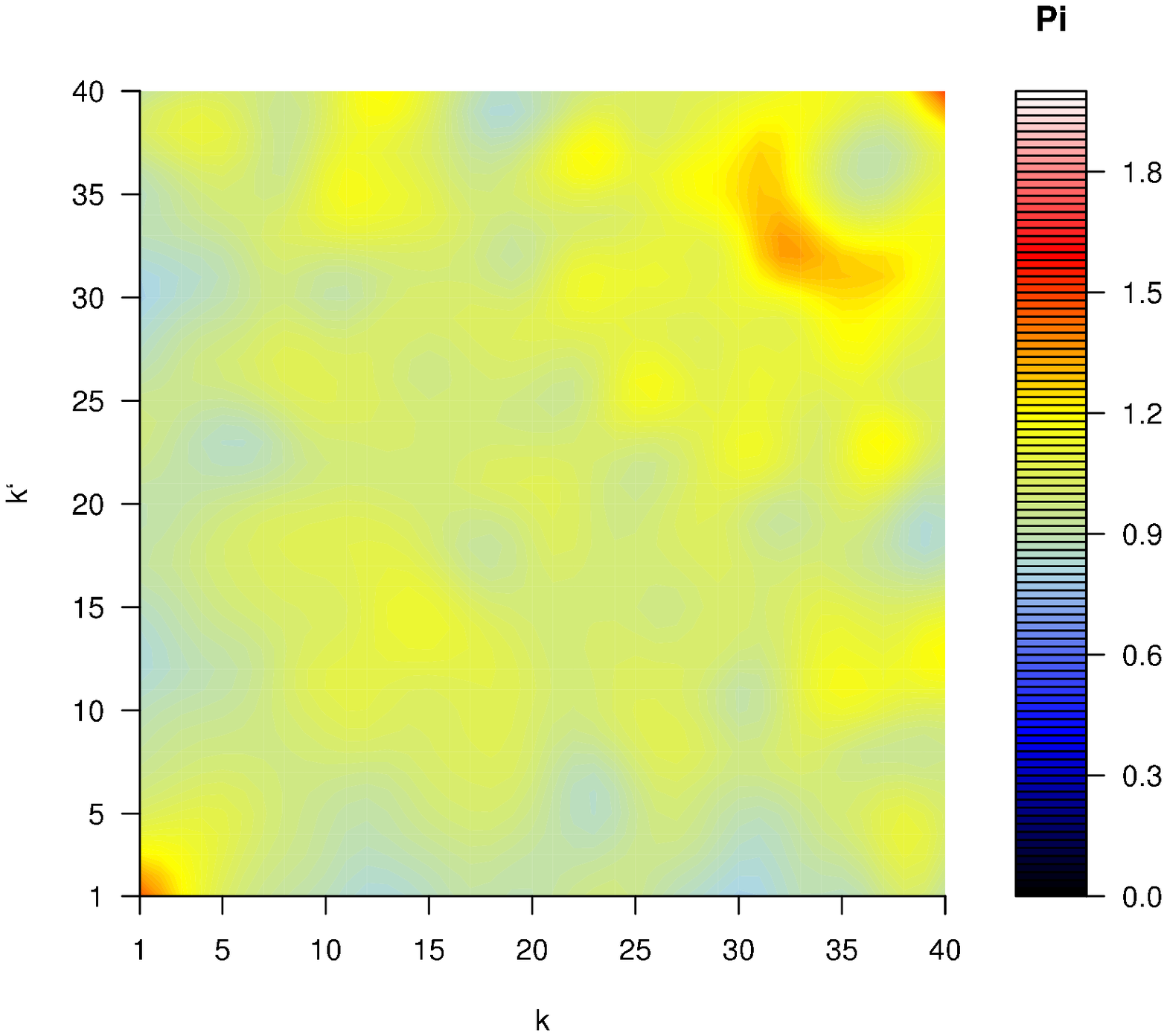}}
\put(120,450){\bf (b)}
\put(0,230){\includegraphics[width=250\unitlength]{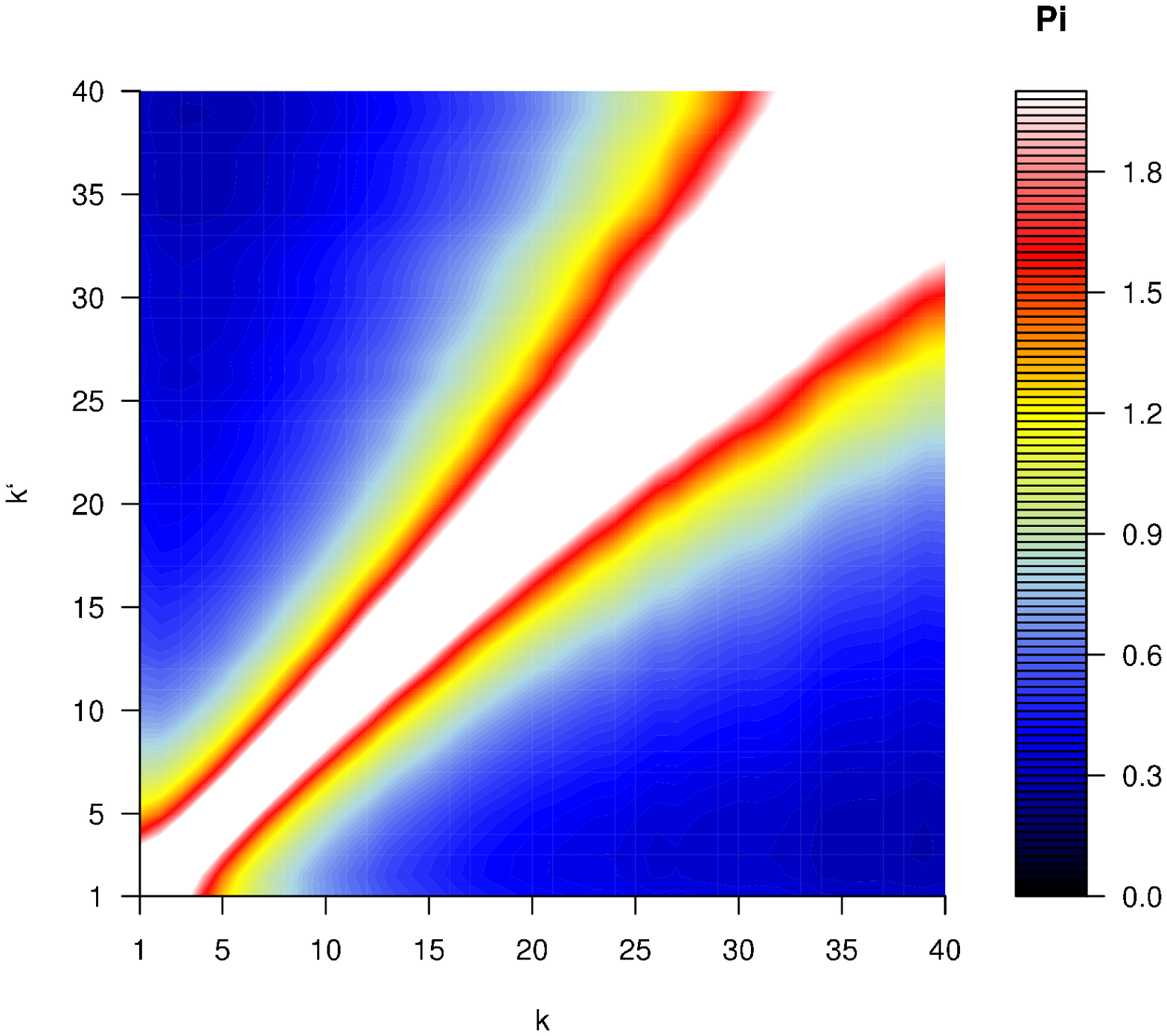}}
\put(410,450){\bf (c)}
\put(290,230){\includegraphics[width=250\unitlength]{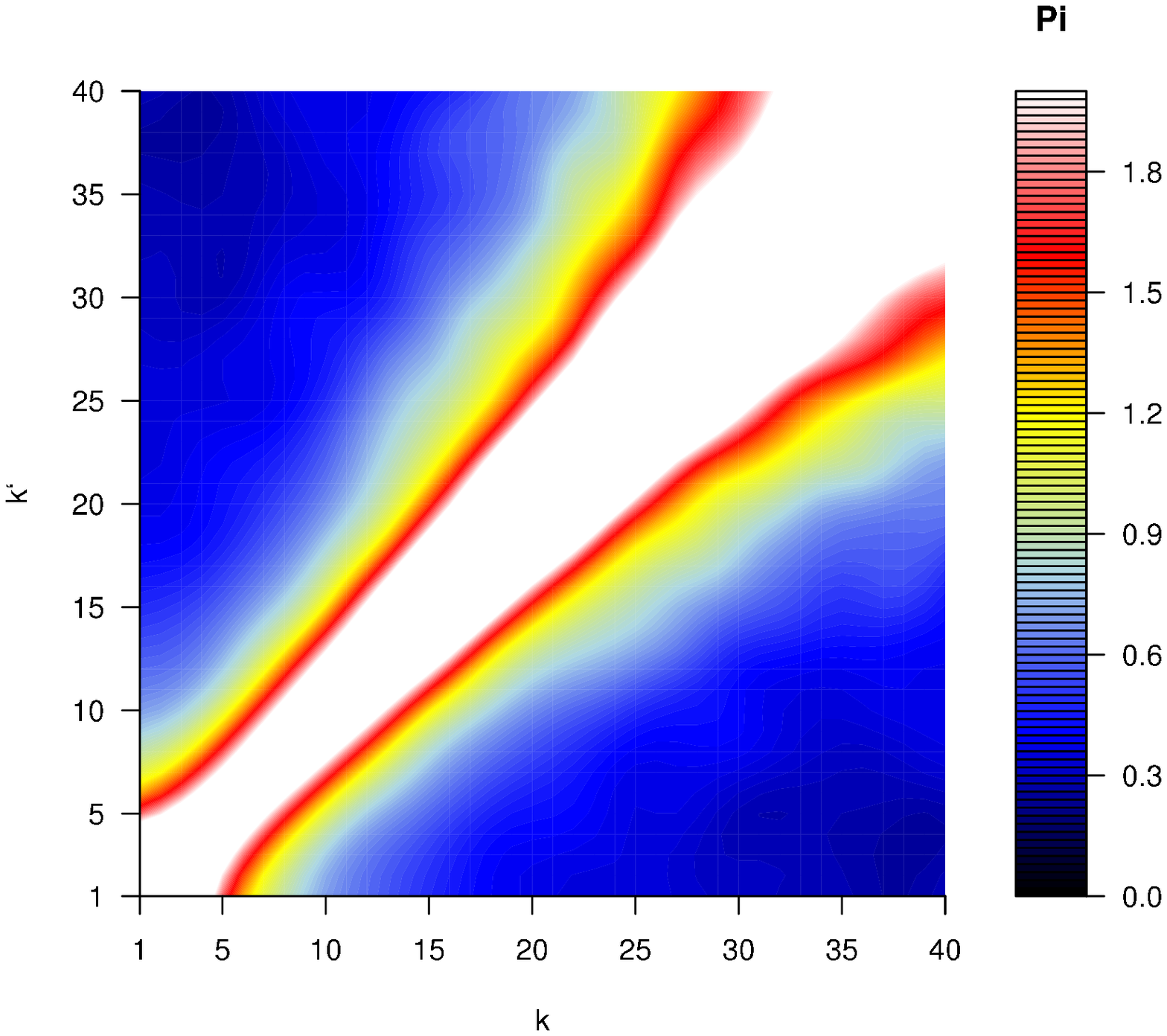}}
\put(120,210){\bf (d)}
\put(0,-10){\includegraphics[width=250\unitlength]{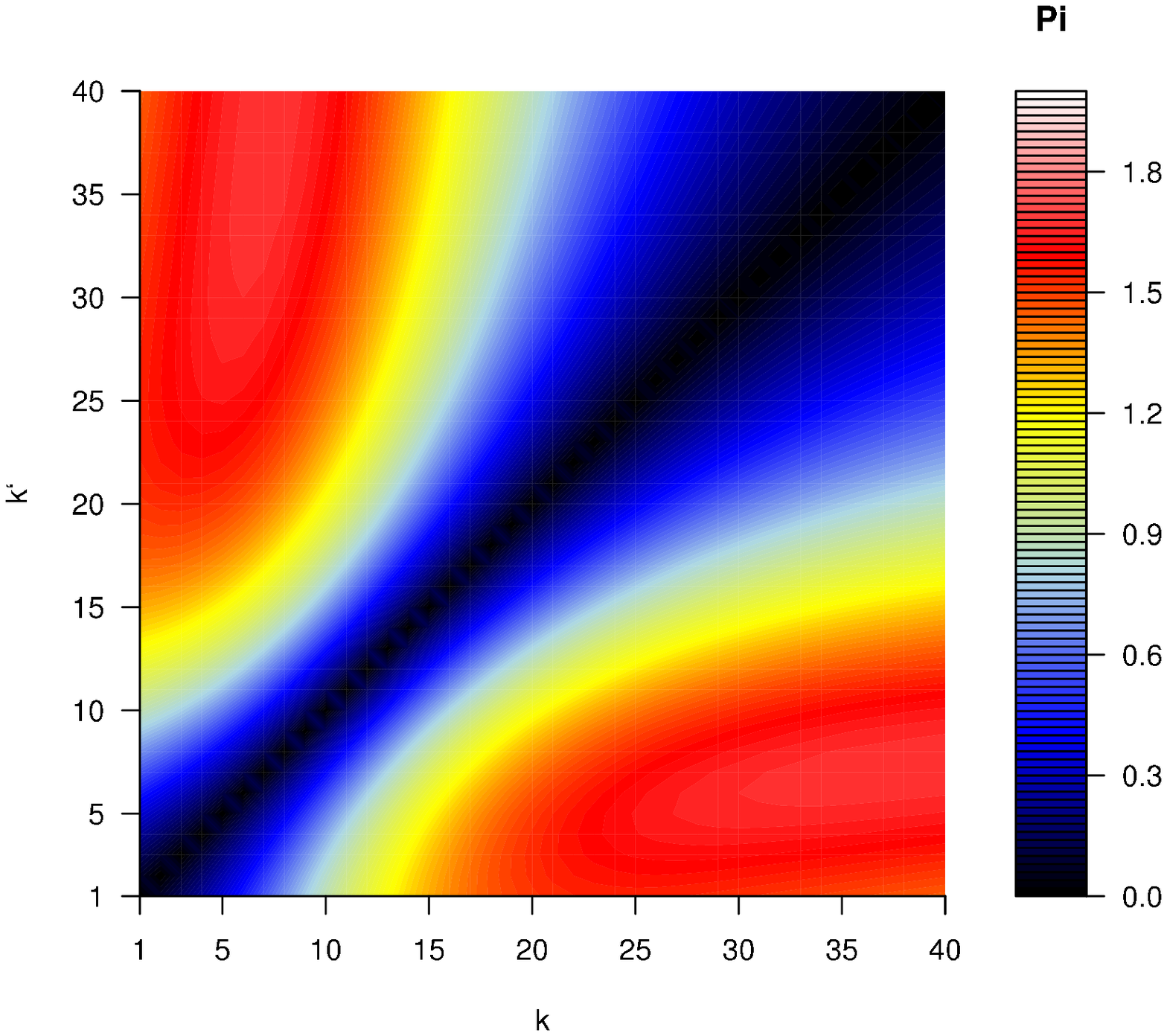}}
\put(410,210){\bf (e)}
\put(290,-10){\includegraphics[width=250\unitlength]{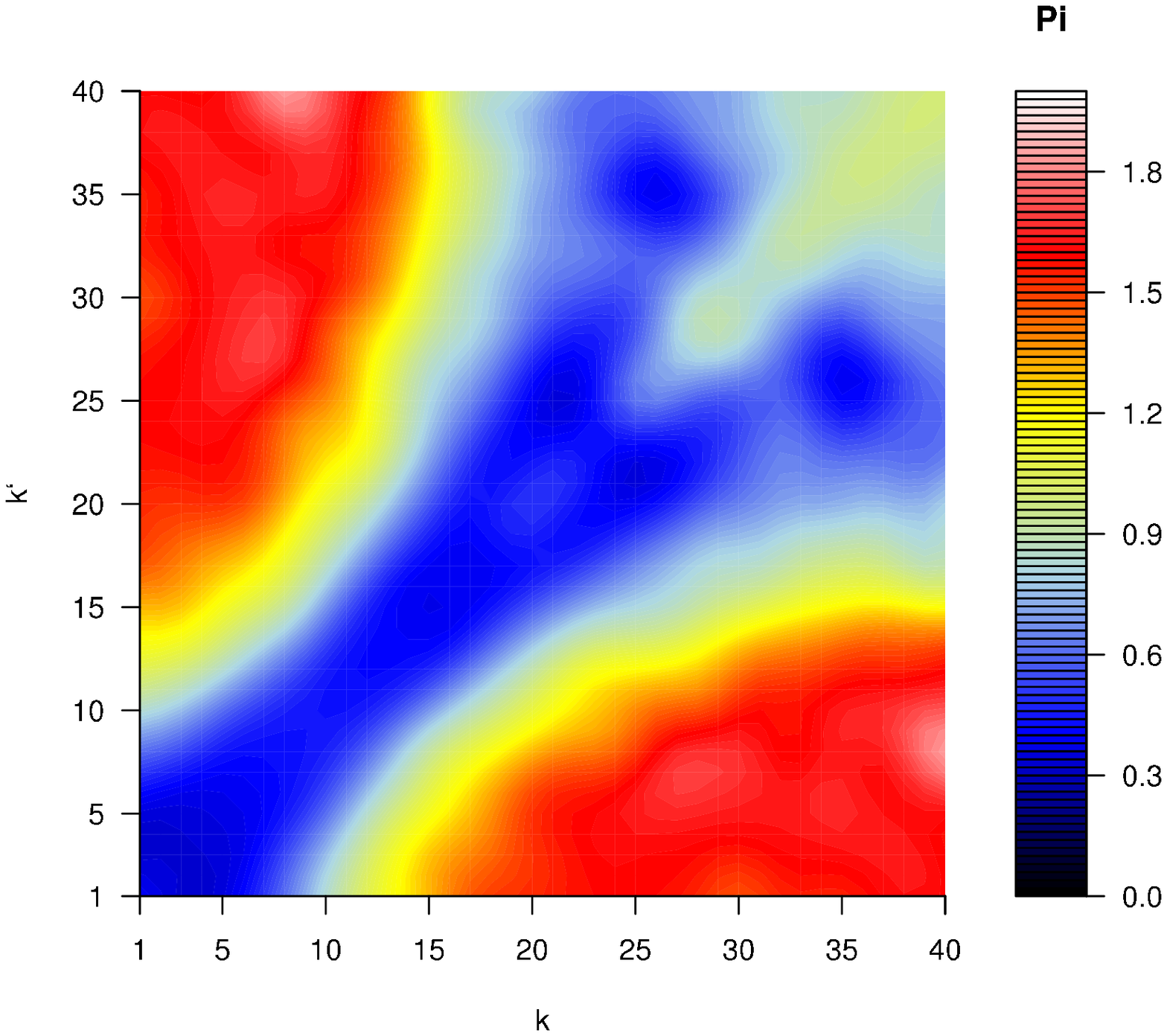}}
\end{picture}
\caption{
Colour plots of the relative degree correlations $\Pi(k,k^\prime)$ of networks which all have the degree sequence
of the {\it Homo sapiens} (from the HPRD database) PPIN (with $N=9463$ and $\bra k\ket=7.4$). (a): $\Pi(k,k^\prime|\bc)$ as measured for the {\it Homo sapiens} PPIN.
(b): the target assortative function $\Pi(k,k^\prime)$ given in (\ref{eq:assortative}). (c): the actual  function  $\Pi(k,k^\prime|\bc^\prime)$ measured after $203,\!441$ accepted moves of the Markov chain in \cite{CooDemAnn09}, on the right. 
(d): the target disassortative function $\Pi(k,k^\prime)$ given in (\ref{eq:disassortative}). (e): the actual  function  $\Pi(k,k^\prime|\bc^\prime)$ measured after $266,\!763$ accepted moves, on the right.
 These results confirm the efficiency of our canonical graph ensemble and its associated Markov chain algorithm, in generating controlled null models.
 }
\label{fig:Pisapiens}
\end{figure}

\section{Discussion}

In this paper we have studied the tailoring of a particular structured random graph ensemble to real-world networks.
We have first derived several mathematical properties of this ensemble, including information-theoretic properties, its Shannon entropy, and the relation between its control parameters and the statistics and correlations of the degrees in the network to which the ensemble is tailored.
We were then able to use the mathematical results in order to derive explicit and  transparent mathematical tools with which to quantify structure in large real networks, define rational distance measures for comparing networks, and for generating controlled null models as benchmark graphs.
These tools are precise and based on information-theoretic principles, yet they take the form of fully explicit formulae (as opposed to
implicit equations that require equilibration of extensive graph simulations). We therefore hope and anticipate that they will be particularly
useful in bio-informatics; indeed a subsequent paper will be fully devoted to their application to a broad range of protein-protein interaction networks, involving multiple organisms and multiple experimental protocols \cite{Fernandes}.

Let us turn to the limitations of this study.
Our work so far has focused on characterizing network structure macroscopically at the level of degree distributions and degree-degree correlations, and was limited to undirected networks and graphs.
We therefore envisage two main directions in which the present theory could and should be developed further. The first, and relatively straightforward, one is generalization of the analysis to tailored {\em directed} random graph ensembles. Here one does not envisage insurmountable obstacles, and it would in bio-informatics open up the possibility of application to e.g. gene regulation networks.
The second direction is towards the inclusion of measures of macroscopic structure that take account of loops, such as the distribution of length-three loops in which individual network nodes participate.
Here the mathematical task is much more challenging, since in entropy calculations it is no longer clear whether and how one can achieve
factorization over nodes.

\section*{Acknowledgements}

ACCC would like to thank Conrad P\'{e}rez-Vicente for stimulating discussions and
the Engineering and Physical Sciences Research Council (UK) for support
in the form of a Springboard Fellowship.

\appendix

\section{\\Degree correlations in the random graph ensemble}
\label{app:Pi_in_Q}

In this appendix we prove the validity of the crucial relation (\ref{eq:Pik}), with $\Pi(k,k^\prime)$ as defined
in (\ref{eq:Pikdef})
for the random graph ensemble (\ref{eq:ensemble}):
\begin{eqnarray}
\hspace*{-10mm}
\Pik
&=&\frac{\kav}{p(k)p(k^\prime)kk^\prime} \lim_{N\to\infty}\sum_{rs}\sum_{\bk}\Big[\prod_\ell p(k_\ell)\Big]\delta_{k,k_r}\delta_{k^\prime,k_s}
\frac{1}{N}\sum_{\bc}\Proba({\bf c}|\bk,Q)c_{rs}
\label{eq:findPi1}
\end{eqnarray}
Let us work out the sum over the graphs $\bc$ in (\ref{eq:findPi1}), using the integral representation $\delta_{nm}=(2\pi)^{-1}\int_0^{2\pi}d\omega~e^{i\omega(n-m)}$ to deal with the $N$ degree constraints $\delta_{k_i,k_i(\bc)}$. This introduces and $N$-fold integration over $\bomega=(\omega_1,\ldots,\omega_N)\in[0,2\pi]^N$. With a modest amount of foresight we introduce the two abbreviations  $\bomega\cdot\bk=\sum_i\omega_ik_i$ and
\begin{eqnarray}
W(\bomega,\bk)&=&
\prod_{i<j}\Big\{1+\frac{\overline{k}
}{N}Q(k_i,k_j)[e^{-i(\omega_i+\omega_j)}\!-\!1]\Big\}
\label{eq:shorthandW}
\end{eqnarray}
These allow us to write
\begin{eqnarray}
\hspace*{-15mm}
\sum_{\bc}\Proba({\bf c}|\bk,Q)c_{rs}
&=&
\frac{\sum_{\bc}c_{rs}
\prod_{i<j}\left[\frac{\overline{k}
}{N}Q(k_i,k_j)\delta_{c_{ij},1}\!+\!\Big(1\!-\!\frac{\overline{k}
}{N}Q(k_i,k_j)\Big)\delta_{c_{ij},0}\right].\prod_i\delta_{k_i,k_i(\bc)}
}
{\sum_{\bc}
\prod_{i<j}\left[\frac{\overline{k}
}{N}Q(k_i,k_j)\delta_{c_{ij},1}\!+\!\Big(1\!-\!\frac{\overline{k}
}{N}Q(k_i,k_j)\Big)\delta_{c_{ij},0}\right].\prod_i\delta_{k_i,k_i(\bc)}
}
\nonumber
\\
\hspace*{-15mm}
&&\hspace*{-25mm}=
\frac{\int\!d\bomega~e^{i\bomega\cdot\bk}\sum_{\bc}c_{rs}
\prod_{i<j}\Big\{\Big[\frac{\overline{k}
}{N}Q(k_i,k_j)\delta_{c_{ij},1}\!+\!\Big(1\!-\!\frac{\overline{k}
}{N}Q(k_i,k_j)\Big)\delta_{c_{ij},0}\Big]e^{-i c_{ij}(\omega_i+\omega_j)}\Big\}
}
{\int\!d\bomega~e^{i\bomega\cdot\bk}\sum_{\bc}
\prod_{i<j}\Big\{\Big[\frac{\overline{k}
}{N}Q(k_i,k_j)\delta_{c_{ij},1}\!+\!\Big(1\!-\!\frac{\overline{k}
}{N}Q(k_i,k_j)\Big)\delta_{c_{ij},0}\Big]e^{-i c_{ij}(\omega_i+\omega_j)}\Big\}
}
\nonumber
\\
\hspace*{-15mm}&=&
\frac{\int\!d\bomega~W(\bomega,\bk) e^{i\bomega\cdot\bk}
\Big[\frac{\frac{\overline{k}
}{N}Q(k_r,k_s)e^{-i (\omega_r+\omega_s)}}
{1+\frac{\overline{k}
}{N}Q(k_r,k_s)[e^{-i (\omega_r+\omega_s)}\!-\!1]}
\Big]
}
{\int\!d\bomega~W(\bomega,\bk) e^{i\bomega\cdot\bk}
}
\nonumber
\\
\hspace*{-15mm}&=&
\frac{\overline{k}}{N}Q(k_r,k_s)[1+\order(N^{-1})]
\frac{\int\!d\bomega~W(\bomega,\bk) e^{i\bomega\cdot\bk-i (\omega_r+\omega_s)}
}
{\int\!d\bomega~W(\bomega,\bk) e^{i\bomega\cdot\bk}
}
\label{eq:Cintermediate1}
\end{eqnarray}
We next expand the function $W(\bomega,\bk)$, as defined in (\ref{eq:shorthandW}), in leading orders for large $N$, using the abbreviation
$P(q,\omega|\bomega,\bk)=N^{-1}\sum_i\delta_{q,k_i}\delta(\omega-\omega_i)$:
\begin{eqnarray}
\hspace*{-15mm}
W(\bomega,\bk)&=&
\prod_{i<j}\exp\Big\{\frac{\overline{k}
}{N}Q(k_i,k_j)[e^{-i(\omega_i+\omega_j)}\!-\!1]+\order(N^{-2})\Big\}
\nonumber
\\
\hspace*{-15mm}
&=& \exp\Big\{\frac{\overline{
k}}{2N}\sum_{ij}Q(k_i,k_j)[e^{-i(\omega_i+\omega_j)}\!-\!1]+\order(1)\Big\}
\nonumber
\\
\hspace*{-15mm}
&=& \exp\Big\{
\frac{1}{2}\overline{k} N\!\sum_{qq^\prime}\!\int\!d\omega d\omega^\prime~P(q,\omega|\bomega,\bk)P(q^\prime\!,\omega^\prime|\bomega,\bk)Q(q,q^\prime)[e^{-i(\omega+\omega^\prime)}\!-\!1]
+\order(1)\Big\}
\nonumber
\\
\hspace*{-15mm}&&
\end{eqnarray}
We now insert the following representation of unity, for each combination of $(q,\omega)$,
\begin{eqnarray}
1&=&\int\!dP(q,\omega)\delta[P(q,\omega)-P(q,\omega|\bomega,\bk)]
\nonumber\\
&=&
\int\!\frac{dP(q,\omega)d\hat{P}(q,\omega)}{2\pi/N}~e^{iN\hat{P}(q,\omega)[P(q,\omega)-P(q,\omega|\bomega,\bk)]}
\end{eqnarray}
and convert the previous expression for $W(\bomega,\bk)$ into the form of a functional integral, with a path integral measure $\{dP\}=\prod_{q,\omega}[dP(q,\omega)\Delta\omega/\sqrt{2\pi}]$ (where the values of $\omega\in[0,2\pi]$ are first discretized, with the discretization spacing $\Delta\omega$ sent to zero as soon as this is possible):
\begin{eqnarray}
W(\bomega,\bk)
&=&\int\{dP d\hat{P}\} e^{iN\sum_q\int\!d\omega~\hat{P}(q,\omega)P(q,\omega)-i\sum_i\hat{P}(k_i,\omega_i)+\order(1)}
\nonumber
\\
&& \times e^{
\frac{1}{2}\overline{k} N\!\sum_{qq^\prime}\!\int\!d\omega d\omega^\prime~P(q,\omega)P(q^\prime\!,\omega^\prime)Q(q,q^\prime)[e^{-i(\omega+\omega^\prime)}-1]
}
\nonumber
\\
&=&\int\{dP d\hat{P}\} e^{N\big(\Psi[\{P,\hat{P}\}]+\Phi[\{P\}]\big)-i\sum_i\hat{P}(k_i,\omega_i)+\order(1)}
\end{eqnarray}
with
\begin{eqnarray}
\Psi[\{P,\hat{P}\}]&=& i\sum_q\int_{0}^{2\pi}\!d\omega~\hat{P}(q,\omega)P(q,\omega)
\label{eq:Psi}
\\
\Phi[\{P\}]&=&
\frac{1}{2}\overline{k}\sum_{qq^\prime}\!\int_{0}^{2\pi}\!d\omega d\omega^\prime~P(q,\omega)P(q^\prime\!,\omega^\prime)Q(q,q^\prime)[e^{-i(\omega+\omega^\prime)}-1]
\label{eq:Phi}
\end{eqnarray}
We can now integrate over the $N$-fold angles $\bomega\in[0,2\pi]^N$, and obtain
\begin{eqnarray}
\hspace*{-15mm}
\int\!d\bomega~W(\bomega,\bk) e^{i\bomega\cdot\bk^\prime}
&=&
\int\{dP d\hat{P}\} e^{N\big(\Psi[\{P,\hat{P}\}]+\Phi[\{P\}]\big)+\order(1)}
\int\!d\bomega~e^{i\bomega\cdot\bk^\prime-i\sum_i\hat{P}(k_i,\omega_i)}
\nonumber
\\
\hspace*{-15mm}
&=&
\int\{dP d\hat{P}\} e^{N\big(\Psi[\{P,\hat{P}\}]+\Phi[\{P\}]\big)+\order(1)}
\prod_i \int\!d\omega~e^{i[\omega k^\prime_i-\hat{P}(k_i,\omega)]}
\end{eqnarray}
and write the ratio of integrals in (\ref{eq:Cintermediate1}) as
\begin{eqnarray}
\hspace*{-25mm}
\frac{\int\!d\bomega~e^{i\bomega\cdot\bk-i(\omega_r+\omega_s)} W(\bomega,\bk)
}
{\int\!d\bomega~e^{i\bomega\cdot\bk} W(\bomega,\bk)
}&=&
\nonumber
\\
\hspace*{-25mm}
&&\hspace*{-41mm}
\frac{\int\{dP d\hat{P}\} e^{N\big(\Psi[\{P,\hat{P}\}]+\Phi[\{P\}]+\Omega[\{\hat{P}\}|\bk]\big)+\order(1)}
\Big\{\frac{\big[\int\!d\omega~ e^{i\omega(k_r-1)-i\hat{P}(k_r,\omega)}\big]\big[\int\!d\omega~ e^{i\omega(k_s-1)-i\hat{P}(k_s,\omega)}\big]}
{\big[\int\!d\omega e^{i\omega~ k_r-i\hat{P}(k_r,\omega)}\big]\big[\int\!d\omega~ e^{i\omega k_s-i\hat{P}(k_s,\omega)}\big]}
\Big\}}
{\int\{dP d\hat{P}\} e^{N\big(\Psi[\{P,\hat{P}\}]+\Phi[\{P\}]+\Omega[\{\hat{P}\}|\bk]\big)+\order(1)}}
\nonumber
\\ \hspace*{-25mm}&&
\end{eqnarray}
with
\begin{eqnarray}
\Omega[\{\hat{P}\}|\bk]&=& \frac{1}{N}\sum_i\log \int\!d\omega~ e^{i[\omega k_i-\hat{P}(k_i,\omega)]}
\end{eqnarray}
Therefore we find upon combining the previous intermediate results that the quantity of interest (\ref{eq:findPi1})
can be written in the following form:
\begin{eqnarray}
\hspace*{-20mm}
\Pik
&=&\frac{\kav^2 Q(k,k^\prime)}{p(k)p(k^\prime)kk^\prime} \lim_{N\to\infty}
\sum_{\bk}\Big[\prod_\ell p(k_\ell)\Big]\Big[\frac{1}{N}\sum_r\delta_{k,k_r}\Big]\Big[\frac{1}{N}\sum_s\delta_{k^\prime,k_s}\Big]\times
\nonumber
\\
\hspace*{-20mm}
&&\hspace*{-13mm}\times
\frac{\int\{dP d\hat{P}\} e^{N\big(\Psi[\{P,\hat{P}\}]+\Phi[\{P\}]+\Omega[\{\hat{P}\}|\bk]\big)+\order(1)}
\Big\{\frac{\big[\int\!d\omega~ e^{i[\omega(k-1)-\hat{P}(k,\omega)]}\big]\big[\int\!d\omega~ e^{i[\omega(k^\prime\!-1)-\hat{P}(k^\prime\!,\omega)]}\big]}
{\big[\int\!d\omega e^{i[\omega~ k-\hat{P}(k,\omega)]}\big]\big[\int\!d\omega~ e^{i[\omega k^\prime\!-\hat{P}(k^\prime\!,\omega)]}\big]}
\Big\}}
{\int\{dP d\hat{P}\} e^{N\big(\Psi[\{P,\hat{P}\}]+\Phi[\{P\}]+\Omega[\{\hat{P}\}|\bk]\big)+\order(1)}}
\nonumber
\\
\hspace*{-20mm}
&=&\frac{\kav^2 Q(k,k^\prime)}{kk^\prime}\times
\nonumber
\\
\hspace*{-20mm}
&&\hspace*{-16mm}\lim_{N\to\infty}
\frac{\int\{dP d\hat{P}\} e^{N\big(\Psi[\{P,\hat{P}\}]+\Phi[\{P\}]+\Omega[\{\hat{P}\}]\big)+\order(1)}
\Big\{\frac{\big[\int\!d\omega~ e^{i[\omega(k-1)-\hat{P}(k,\omega)]}\big]\big[\int\!d\omega~ e^{i[\omega(k^\prime\!-1)-\hat{P}(k^\prime\!,\omega)]}\big]}
{\big[\int\!d\omega e^{i[\omega~ k-\hat{P}(k,\omega)]}\big]\big[\int\!d\omega~ e^{i[\omega k^\prime\!-\hat{P}(k^\prime\!,\omega)]}\big]}
\Big\}}
{\int\{dP d\hat{P}\} e^{N\big(\Psi[\{P,\hat{P}\}]+\Phi[\{P\}]+\Omega[\{\hat{P}\}]\big)+\order(1)}}
\nonumber
\\
\hspace*{-20mm}&&\label{eq:Cintermediate2}
\end{eqnarray}
where
\begin{eqnarray}
\Omega[\{\hat{P}\}]&=& \sum_{k^\pprime}p(k^\pprime)\log \int\!d\omega~ e^{i[\omega k^\pprime-\hat{P}(k^\pprime\!,\omega)]}
\label{eq:Omega}
\end{eqnarray}
We conclude from (\ref{eq:Cintermediate2}), in which the functional integrals can be done by steepest descent in the limit $N\to\infty$,  that $\Pi(k,k^\prime)$ takes the form
\begin{eqnarray}
\Pi(k,k^\prime)&=& Q(k,k^\prime)/F(k|Q)F(k^\prime|Q)
\label{eq:result_of_appendixC}
\end{eqnarray}
with
\begin{eqnarray}
\frac{1}{F(k|Q)}&=&\frac{\bra k\ket}{k}\frac{\int\!d\omega ~e^{i\omega(k-1)-i\hat{P}(k,\omega)}}
{\int\!d\omega ~e^{i\omega k-i\hat{P}(k,\omega)}}
\label{eq:f1}
\end{eqnarray}
and where the functions $P(k,\omega)$ and $\hat{P}(k,\omega)$ are to be solved from extremization of $\Psi[\{P,\hat{P}\}]+\Phi[\{P\}]+\Omega[\{\hat{P}\}]$,
with the three functions given in (\ref{eq:Psi},\ref{eq:Phi},\ref{eq:Omega}), leading to the two coupled functional saddle-point equations
$\delta[\Psi+\Phi]/\delta P=0$ and $\delta[\Psi+\Omega]/\delta \hat{P}=0$.

The last step in this appendix is to derive from the saddle-point equations an equation for the function $F(k|Q)$ in
(\ref{eq:result_of_appendixC}).
Upon transforming $\exp[-i\hat{P}(k,\omega)]=R(k,\omega)$, our saddle-point equations simplify to
\begin{eqnarray}
R(k,\omega)
&=&
\exp\Big\{\bra k\ket\!\sum_{k^\prime}\!\int\!d\omega^\prime~P(k^\prime\!,\omega^\prime)Q(k,k^\prime)[e^{-i(\omega+\omega^\prime)}\!-\!1]
\Big\}
\\
P(k,\omega)
&=&
 p(k)
 \frac{ R(k,\omega)e^{i\omega k}}
{\int\!d\omega^\prime~ R(k,\omega^\prime)e^{i\omega^\prime k}}
\end{eqnarray}
Elimination of $P(k,\omega)$ from this set gives, using the identity $\int\!d\omega~P(k,\omega)=p(k)$,
\begin{eqnarray}
R(k,\omega)
&=&
\exp\left\{\bra k\ket\!\sum_{k^\prime}p(k^\prime)Q(k,k^\prime)\left[e^{-i\omega}
 \frac{\int\!d\omega^\prime~ R(k^\prime,\omega^\prime)e^{i\omega^\prime (k^\prime-1)}}
{\int\!d\omega^\prime~ R(k^\prime,\omega^\prime)e^{i\omega^\prime k^\prime}}
-1\right]
\right\}
\nonumber
\\
&=&
\exp\Big\{\sum_{k^\prime}p(k^\prime)Q(k,k^\prime)e^{-i\omega}k^\prime /F(k^\prime|Q)
-G(k|Q)\Big\}
\end{eqnarray}
in which $F(k|Q)$ is defined in  (\ref{eq:f1}), and
$G(k|Q)= \bra k\ket\!\sum_{k^\prime}p(k^\prime)Q(k,k^\prime)$.
Insertion of our expression for $R(k,\omega)$ into (\ref{eq:f1}), using $\exp[-i\hat{P}(k,\omega)]=R(k,\omega)$, leaves us with an equation for $F(k|Q)$ only, from which the object $G(k|Q)$ simply drops out since it gives an identical pre-factor $\exp[-G(k|Q)]$ in both the  numerator and the denominator of the formula for $F(k|Q)$:
\begin{eqnarray}
\frac{1}{F(k|Q)}&=&\frac{\bra k\ket}{k}\frac{\int\!d\omega ~e^{i\omega(k-1)}R(k,\omega)}
{\int\!d\omega ~e^{i\omega k}R(k,\omega)}
\nonumber
\\
&=&\frac{\bra k\ket}{k} \frac{\sum_{m\geq 0}\frac{1}{m!}\Big[\sum_{k^\prime}p(k^\prime)Q(k,k^\prime)k^\prime/F(k^\prime|Q)\Big]^m
\int\!d\omega ~e^{i\omega(k-1)-im\omega}}
{\sum_{m\geq 0}\frac{1}{m!}\Big[\sum_{k^\prime}p(k^\prime)Q(k,k^\prime)k^\prime F(k^\prime|Q)\Big]^m
\int\!d\omega ~e^{i\omega k-im\omega}}
\nonumber
\\
&=& \frac{\bra k\ket}{k}\frac{\sum_{m\geq 0}\frac{1}{m!}\delta_{m,k-1}\Big[\sum_{k^\prime}p(k^\prime)Q(k,k^\prime)k^\prime F(k^\prime|Q)\Big]^m}
{\sum_{m\geq 0}\frac{1}{m!}\delta_{mk}\Big[\sum_{k^\prime}p(k^\prime)Q(k,k^\prime)k^\prime/F(k^\prime|Q)\Big]^m}
\nonumber
\\
&=& \frac{\bra k\ket}
{\sum_{k^\prime}p(k^\prime)Q(k,k^\prime)k^\prime /F(k^\prime|Q)}
\end{eqnarray}
Equivalently:
\begin{eqnarray}
F(k|Q)= \bra k\ket^{-1}\sum_{k^\prime}p(k^\prime)k^\prime Q(k,k^\prime) F^{-1}(k^\prime|Q)
\label{eq:resultforF}
\end{eqnarray}
Note that the present derivation of the combined result (\ref{eq:result_of_appendixC},\ref{eq:resultforF})
also serves as the explicit proof of the validity of corrigendum \cite{PerCoo09}.

\end{document}